\documentclass[onecolumn]{emulateapj}
\usepackage{amssymb,amsmath,xcolor}
\usepackage[T2A,T1]{fontenc}
\usepackage[utf8]{inputenc}
\usepackage[russian,english]{babel}
\usepackage[warn]{textcomp}
\newcommand*{\No}{\textnumero}

\newcommand{\pam}{\textsf{PAMELA}}
\newcommand{\deu}{\textsuperscript{2}H}
\newcommand{\prot}{\textsuperscript{1}H}
\newcommand{\het}{\textsuperscript{3}He}
\newcommand{\hef}{\textsuperscript{4}He}
\newcommand{\lisi}{\textsuperscript{6}Li}
\newcommand{\lise}{\textsuperscript{7}Li}
\newcommand{\bese}{\textsuperscript{7}Be}
\newcommand{\beni}{\textsuperscript{9}Be}
\newcommand{\bete}{\textsuperscript{10}Be}

\submitted{}

\begin{document}

\title{Lithium and Beryllium isotopes with the \pam\ experiment}

  \author{
W. Menn$^{16}$,	E. A. Bogomolov$^{9}$, 
	M. Simon$^{16}$, G. Vasilyev$^{9}$,\\
O. Adriani$^{1,2}$, G. C. Barbarino$^{3,4}$,
  G. A. Bazilevskaya$^{5}$, R. Bellotti$^{6,7}$, 
M. Boezio$^{8}$, 
 M. Bongi$^{1,2}$,\\
 V. Bonvicini$^{8}$,
S. Bottai$^{2}$, A. Bruno$^{19,7}$, F. Cafagna$^{7}$,
D. Campana$^{4}$,
P. Carlson$^{13}$, 
M. Casolino$^{11}$,\\
 G. Castellini$^{14}$, C.~De~Donato$^{10}$, 
C. De Santis$^{11}$, N. De
Simone$^{10}$, V. Di Felice$^{10}$, V. Formato$^{8,\dagger}$,\\ 
 A. M. Galper$^{12}$,
A. V. Karelin$^{12}$, 
S. V. Koldashov$^{12}$, S. Koldobskiy$^{12}$, S. Y. Krutkov$^{9}$,  
A. N. Kvashnin$^{5}$,\\ A. Leonov$^{12}$, 
V. Malakhov$^{12}$, L. Marcelli$^{11}$,  M. Martucci$^{11,15}$, 
A. G. Mayorov$^{12}$,  
M Merg\`{e}$^{10,11}$,\\
V. V. Mikhailov$^{12}$,
E. Mocchiutti$^{8}$, 
A. Monaco$^{6,7}$,  
N. Mori$^{2}$, R.~Munini$^{8}$,
G. Osteria$^{4}$,
F. Palma$^{10,11}$,\\
B.~Panico$^{4}$,
P. Papini$^{2}$,
M. Pearce$^{13}$, P. Picozza$^{10,11}$, 
M. Ricci$^{15}$,
S. B. Ricciarini$^{14}$,\\ 
R. Sarkar$^{17,*}$, V. Scotti$^{3,4}$, 
 R. Sparvoli$^{10,11}$, P. Spillantini$^{12,20}$, 
Y. I. Stozhkov$^{5}$, A. Vacchi$^{8,18}$, \\
E. Vannuccini$^{2}$,  S. A. Voronov$^{12}$,
 Y. T. Yurkin$^{12}$, 
 G. Zampa$^{8}$, N. Zampa$^{8}$}
\affil{$^{1}$University of Florence, Department of Physics, I-50019 Sesto Fiorentino, Florence, Italy}
\affil{$^{2}$INFN, Sezione di Florence, I-50019 Sesto Fiorentino, Florence, Italy}
\affil{$^{3}$University of Naples ``Federico II'', Department of Physics, I-80126 Naples, Italy}
\affil{$^{4}$INFN, Sezione di Naples,  I-80126 Naples, Italy}
\affil{$^{5}$Lebedev Physical Institute, RU-119991, Moscow, Russia}
\affil{$^{6}$University of Bari, Department of Physics, I-70126 Bari, Italy}
\affil{$^{7}$INFN, Sezione di Bari, I-70126 Bari, Italy}
\affil{$^{8}$INFN, Sezione di Trieste, I-34149 Trieste, Italy}
\affil{$^{9}$Ioffe Physical Technical Institute,  RU-194021 St. Petersburg, Russia}
\affil{$^{10}$INFN, Sezione di Rome ``Tor Vergata'', I-00133 Rome, Italy}
\affil{$^{11}$University of Rome ``Tor Vergata'', Department of Physics,  I-00133 Rome, Italy}
\affil{$^{12}$National Research Nuclear University MEPhI, RU-115409 Moscow}
\affil{$^{13}$KTH, Department of Physics, and the Oskar Klein Centre for Cosmoparticle Physics, AlbaNova University Centre, SE-10691 Stockholm, Sweden}
\affil{$^{14}$IFAC, I-50019 Sesto Fiorentino, Florence, Italy}
\affil{$^{15}$INFN, Laboratori Nazionali di Frascati, Via Enrico Fermi 40, I-00044 Frascati, Italy}
\affil{$^{16}$Universit\"{a}t Siegen, Department of Physics, D-57068 Siegen, Germany}
\affil{$^{17}$Indian Centre for Space Physics, 43 Chalantika, Garia Station Road, Kolkata 700084, West Bengal, India}
\affil{$^{18}$University of Udine, Department of Mathematics and Informatics, I-33100 Udine, Italy}
\affil{$^{19}$Heliophysics Division, NASA Goddard Space Flight Center, Greenbelt, MD, USA}
\affil{$^{20}$APS/INAF, I-00133 Rome, Italy}
\affil{$^{\dagger}$Now at INFN, Sezione di Perugia, I-06123 Perugia, Italy}
\affil{$^{*}$Previously at INFN, Sezione di Trieste, I-34149 Trieste, Italy}

\begin{abstract}
The cosmic-ray lithium and beryllium (\lisi,\lise,\bese,\beni,\bete) isotopic composition has been measured with the  satellite-borne experiment \pam, which was launched into low-Earth orbit on-board the Resurs-DK1 satellite on June 15\textsuperscript{th} 2006.
The rare lithium and beryllium isotopes in cosmic rays are believed to originate mainly from the interaction of high energy carbon, nitrogen and oxygen nuclei with the interstellar medium (ISM), but also on ``tertiary'' interactions in the ISM (i.e. produced by further fragmentation of secondary beryllium and boron).
In this paper the isotopic ratios \lise/\lisi\ and \bese/(\beni+\bete) measured between 150 and 1100 MeV/n using two different detector systems from July 2006 to September 2014 will be presented.
\end{abstract}


\section{Introduction}

Measurements of the spectra or the isotopic composition of elements of the cosmic radiation provide significant constraints on cosmic ray (CR) source composition and CR transport and acceleration in the Galaxy, with an important role played by the secondary-over-primary ratios of nuclear species.

While the primary nuclides are almost exclusively accelerated in astrophysical sources and then injected into the interstellar medium (ISM), the secondary nuclides result from inelastic interactions of heavier CR nuclei with the ISM. 
The relative abundance of secondary to primary nuclei is uniquely related to propagation processes and can be used to constrain the models, provided that cross sections and decay chains for all the relevant nuclear processes are known.

Together with \deu\ and \het, lithium, beryllium and boron in cosmic rays are the lightest and most abundant group of elements of almost pure secondary origin.
Among them, boron is mainly produced by fragmentation of carbon, which originates almost entirely from the acceleration sites. The boron to carbon flux ratio has been widely studied and it is considered as the ``standard tool'' for studying propagation models and to determine the key parameters of the models. \citep{2007ARNPS..57..285S}.

It is furthermore important to test the ``universality'' of CR propagation with nuclei of different mass-to-charge ratios: since \deu\ and \het\ CRs are mainly produced by the breakup of the primary \hef\ in the Galaxy, the ratios \deu/\hef\ and \het/\hef\ probe the propagation history of helium rather then carbon \citep{Webber1997}.

If the CR propagation models are tuned on the secondary to primary ratios, they must correctly reproduce the secondary to secondary ratios as well.
Thus ratios such as \deu/\het, \lise/\lisi\ and \bese/(\beni+\bete), which are less sensitive to the astrophysical aspects of a given propagation model, provide a useful consistency check for the calculations.

While the main contribution to the lithium and beryllium isotopes comes from spallation reactions between heavier CR and the ISM, there is also a non-negligible fraction of tertiary origin, i.e. produced by further fragmentation of secondary beryllium and boron. 

The \lisi\ isotope is a pure product of interactions of galactic CR with the ISM, while it is known that the \lise\ isotope has additional sources, like a stellar production and primordial nucleosynthesis \citep{Reeves1994}.

\bese\ decays only by electron capture (half-life on Earth = 53 days), thus the half-life in space depends on the electron density and the galactic cosmic ray lifetime. \beni\ is stable, but \bete\ has a half-life of 1.5 x 10$^6$ years, which is comparable to the characteristic storage time expected for the galactic containment.

These characteristics makes the knowledge of the abundance of the lithium and beryllium isotopes a complementary tool to tune propagation models, helping in removing parameter degeneracy and giving a more detailed description of the galactic propagation process. On the other hand, the interpretation of lithium and beryllium abundances requires the knowledge of a complex chain of nuclear reactions, whose cross sections are still affected by large uncertainty \citep{Tomassetti2012}.

In this paper results of the \pam\ satellite experiment are presented. It was mounted on the Resurs DK1 satellite and launched from the Baykonur cosmodrome on June 15th 2006. \pam\ was put in a polar elliptical orbit at an altitude between $\sim 350$ and $\sim 600$ km with an inclination of $70^\circ$. After 2010 the orbit was changed and became circular at a constant altitude of 570 km.

The \pam\ measurements of the boron and carbon spectra and ratio as well as the fluxes and ratios of the hydrogen and helium isotopes \prot,\deu,\het,\hef\ have already been published
\citep{2014ApJ...791..93}, \citep{2016ApJ...818..1}.

For this analysis of the lithium and beryllium isotopes data gathered between July 2006 and September 2014 was used. About 60000 lithium nuclei and about 32000 beryllium nuclei were selected in the energy interval between 100 and 2000 MeV/n.
\\

\section{The \pam\ instrument}
\begin{figure}[t]
    \centering
    \epsscale{0.5}
    \plotone{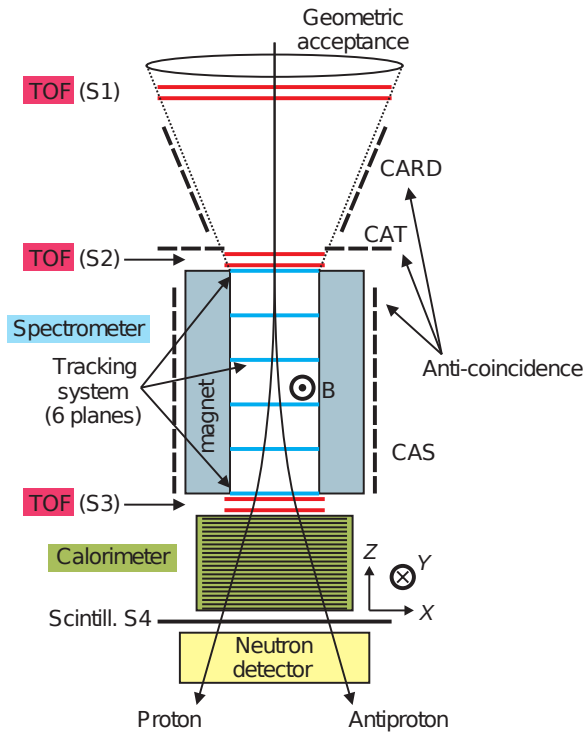}
    \caption{Scheme of the detectors composing the \pam\ satellite experiment.}
    \label{im:pamela}
\end{figure}

The \pam\ satellite-borne cosmic ray experiment was built to measure charged particles in the cosmic radiation with a particular focus on antiparticles. 
The \pam\ apparatus is composed of several sub-detectors: Time-of-Flight (ToF) system, anti-coincidence system (CARD, CAS, CAT), magnetic spectrometer with microstrip silicon tracking system, W/Si electromagnetic imaging calorimeter, shower-tail-catcher scintillator (S4) and neutron detector. The apparatus is schematically shown in Fig.~\ref{im:pamela}

For a detailed description of the \pam\ instrument and an overview of the mission see \cite{2007APh....27..296P}, \cite{NuovoCimento2017}.
The core of the instrument is a magnetic spectrometer, made of a permanent magnet (0.43 T) and a silicon tracking system (resolution in the bending side 4 $\mu$m) for a maximum detectable rigidity of 1 TV.  The momentum resolution for lithium and beryllium nuclei is better than 4\% between 2 GV and 20 GV. 
The ToF system consists of three double layers (S1, S2, S3) of segmented plastic scintillator strips, placed in alternate layers orthogonal to each other. The first plane (S1) is placed on top of the instrument, the second (S2) and third (S3) planes are placed above and below the magnetic cavity. The layers of S1 and S3 are 7 mm thick, while those of S2 are only 5 mm thick.
Making use of different combinations of layers, the ToF system can provide up to 12 measurements of the particle velocity $\beta = v/c$. A weighted mean technique is then employed to derive an overall value for $\beta$ from these measurements  . 
The overall time resolution of the ToF system is about 250 ps for $Z = 1$ particles, about 100 ps for $Z = 2$ particles \citep{2016ApJ...818..1}, about 85 ps for $Z = 3$ particles and about 80 ps for $Z = 4$ particles.  

The W/Si sampling imaging calorimeter comprises 44 single-sided silicon strip detector planes interleaved with 22 plates of tungsten absorber \citep{2002NIMPA.487..407B}. Each tungsten layer has a thickness of 0.74 radiation lengths (2.6 mm) and it is sandwiched between two printed circuit boards, which house the silicon detectors as well as the front-end and digitizing electronics. Each silicon plane consists of 3x3, 380 $\mu$m thick, 8x8 cm$^2$ detectors, segmented into 32 strips with a pitch of 2.4 mm. The orientation of the strips for two consecutive silicon planes is shifted by 90 degrees, thus providing 2-dimensional spatial information. The total depth of the calorimeter is 16.3 radiation lengths and 0.6 nuclear interaction lengths.

\section{Data analysis}
\subsection{Event selection}   \label{sec:event_selection}

The selection criteria applied to each triggered event did select positively charged particles with a precise measurement of the absolute value of the particle rigidity and velocity. 
The analysis procedure was quite similar to the previous work on the hydrogen and helium isotopes \citep{2016ApJ...818..1}, a few selections had to be modified due to the higher energy loss of lithium and beryllium. Events were selected requiring:
\begin{itemize}
\item A single track fitted within the spectrometer fiducial volume where the reconstructed track is at least 1.5 mm away from the magnet walls. This analysis employs the position finding algorithm used for the analysis of the boron and carbon spectra \citep{2014ApJ...791..93} which was developed to account for the ionization energy losses that might saturate the read-out electronics for the silicon layers of the tracking system. 

\item A positive value for the reconstructed track curvature and a positive value for the measured time of flight. This selection ensures that the particle enters \pam\ from above.

\item Selected tracks must have at least 3 hits on the $x$-view and at least 2 hits on the $y$-view, for the $x$-view a ``lever arm'' (distance between the lowermost and uppermost layer) $\geq 4$ was required. 
To increase the statistics these selection criteria are less strict then the ones used in the work on the hydrogen and helium isotopes, where at least 4 hits on the $x$-view (bending view) and at least 3 hits on the $y$-view were required \citep{2016ApJ...818..1}. 
The full Monte Carlo simulation of the \pam\ apparatus, based on the \texttt{GEANT4} code \citep{Geant4} and described in \cite{2013ApJ...770..2}, was used to study the performance of these relaxed tracking criteria. In the rigidity range of this analysis (1 - 5 GV) no measurable effect was found.

\item Charge selection: Hydrogen and helium events were identified using the ionization measurements provided by the silicon sensors of the magnetic spectrometer \citep{2016ApJ...818..1} and rejected from the data sample. In the following lithium and beryllium events were selected by means of ionization energy losses in the ToF system. Each of the 48 channels has been calibrated for a conversion of the $dE/dx$ to a charge value by using the velocity information $\beta$ from the ToF. The charge for one paddle is then calculated by taking the arithmetic mean of the two PMTs. 
Finally charge consistency has been required between S12 (the lower layer of the two layers constituting S1) and $\langle$S2$\rangle$ and $\langle$S3$\rangle$ (the arithmetic mean of the ionization for the two layers constituting S2 and S3, respectively). In Fig.~\ref{im:charge_sel} the charge for S12, $\langle$S2$\rangle$, and $\langle$S3$\rangle$ as a function of the rigidity is shown. The actual selection of $Z=3$ or $Z=4$ particles is depicted by the solid lines. S11, the upper layer of the two layers constituting S1, was used for efficiency measurements section, see section \ref{sec:isotopic_ratios}.

\begin{figure}[t]
 \centering
  \plotone{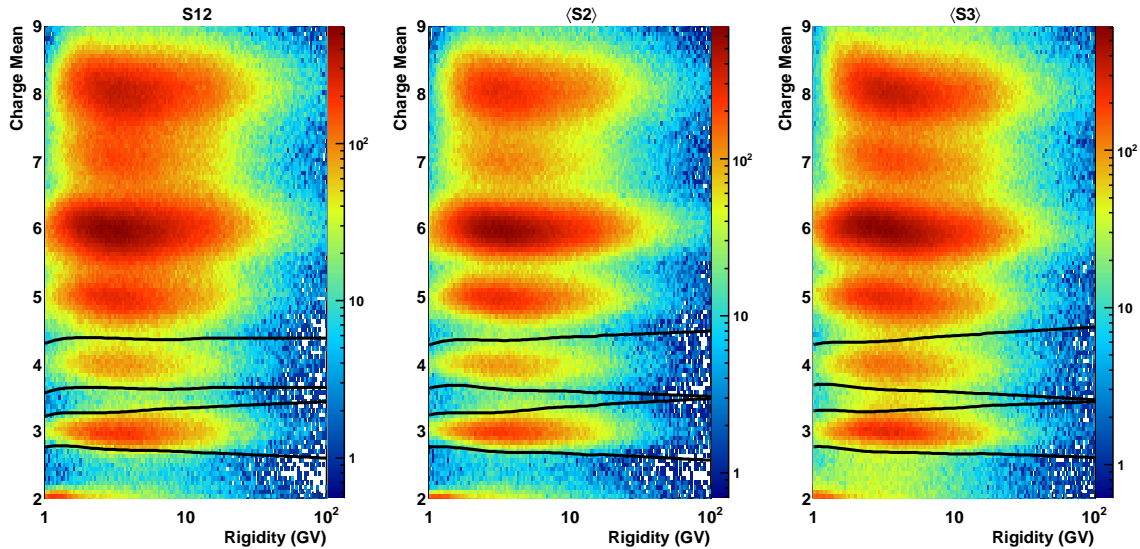}
  \caption{Charge for S12, $\langle$S2$\rangle$, and $\langle$S3$\rangle$ as a function of the rigidity. The actual selection of $Z=3$ or $Z=4$ particles is depicted by the solid lines}
\label{im:charge_sel}
\end{figure}

Note that since $\beta$ is used to derive the charge, doing the selection as a function of rigidity as shown in Fig.~\ref{im:charge_sel} has no effect on the isotopic fluxes. Requiring charge consistency above
and below the tracking system rejected events interacting in the silicon layers \citep{2014ApJ...791..93}. 

\item Galactic events were selected by imposing that the lower edge of the rigidity bin to which the event belongs exceeds the critical rigidity, $\rho_c$, defined as 1.3 times the cutoff rigidity $\rho_{SVC}$ computed in the St\"{o}rmer vertical approximation. 

\end{itemize}

\subsection{Isotope separation in the \pam\ instrument}
Analogue to the analysis described in \citep{2016ApJ...818..1} an isotopic separation at fixed rigidity for each sample of $Z=3$ and $Z=4$ particles is performed. This is possible since the mass of each particle follows the relation $ m = RZe / {\gamma\beta c} $ ($R$ is the magnetic rigidity, $Z \times e$ is the particle charge, and $\gamma$ is the Lorentz factor). The particle velocity $\beta$ is provided either directly from the timing measurement of the ToF system, or indirectly from the energy loss in the calorimeter, which follows $\beta$ via the
Bethe-Bloch formula $dE/dx \propto {Z^2}/{\beta^2}$ (neglecting logarithmic terms).

\begin{figure}[t]
    \centering
    \plotone{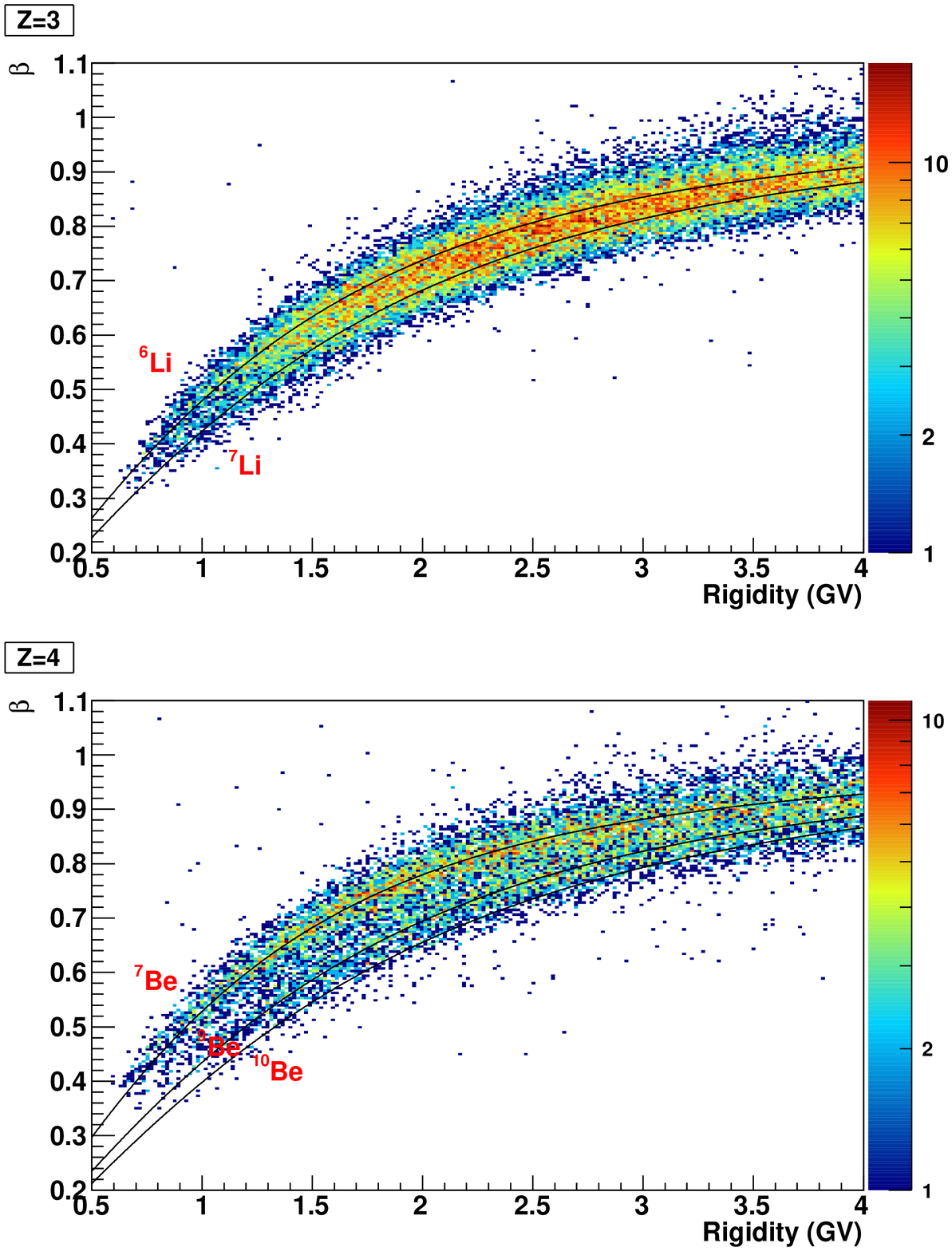}
  \caption{ $\beta$ vs. rigidity for $Z=3$ (\emph{top}) and $Z=4$ (\emph{bottom}) particles. The black lines represent the expectations for each isotope.}
\label{im:beta_r}
\end{figure}

For the ToF analysis the $\beta$ provided by the timing measurement was used. In Fig.~\ref{im:beta_r} $\beta$ vs. the particle rigidity for $Z = 3$ and $Z = 4$ data is shown, the black lines in the figure represent the expectations for each isotope. 

\begin{figure}[t]
    \centering
    \plotone{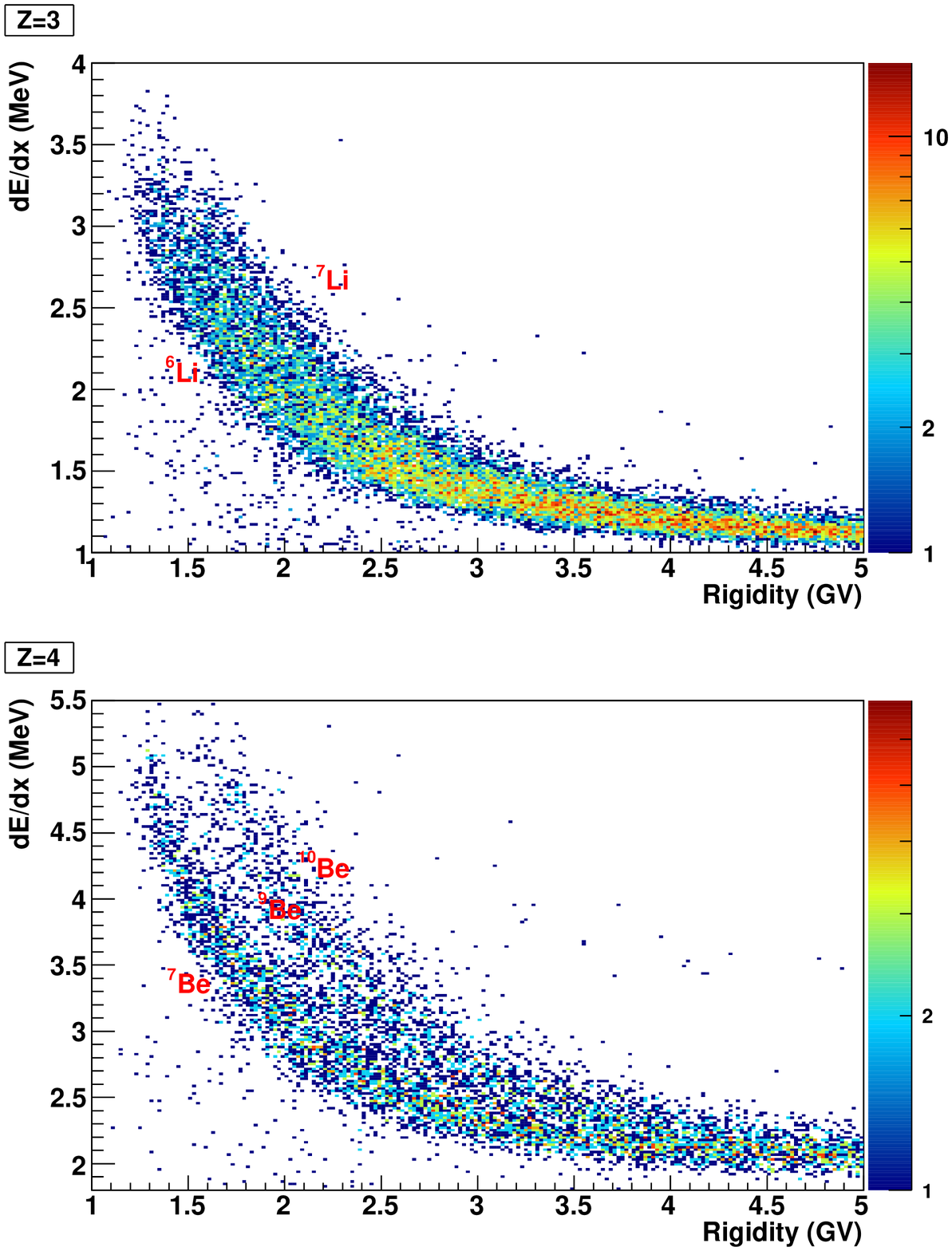}
\caption{Isotopic separation for $Z=3$ (\emph{top}) and $Z=4$ (\emph{bottom}) particles using the truncated mean of the calorimeter $dE/dx$. }
\label{im:calo_tm_r}
\end{figure}

The isotopic analysis of nuclei with the calorimeter is restricted to non-interacting events inside the calorimeter. To check if an interaction occurs, events were selected by applying cuts on the ratio $\Sigma q_{track}/ \Sigma q_{tot}$ in each layer, where $q_{track}$ is the energy deposited in the strip closest to the track and the neighboring strip on each side, and $q_{tot}$ the total energy detected in that layer. In case of no interaction, the fraction of $q_{track}/q_{tot}$ will be equal to one, if strips outside the track are hit, the value is less then one. 
The selection used in this work is the same as described in more detail in \cite{2016ApJ...818..1}: Starting from the top of the calorimeter, $q_{track}/q_{tot}$ is calculated  at each layer, as long as this value is greater than 0.9, these layers are used for further analysis. A value of 0.9 was chosen since it was found to give a good compromise between high efficiency and rejection of interactions.
The $dE/dx$ measurements passing this selection are then sorted and the 50\% of samples with larger pulse amplitudes are excluded before taking the mean of the remaining measurements. This truncation method improves the quality of the mean $dE/dx$ value, since in each silicon layer the energy loss distributions shows a Landau tail, which degrades the resolution of the $dE/dx$ measurement.
Finally an energy dependent lower limit on the remaining number of layers was used, the selection criteria are the same as in \cite{2016ApJ...818..1} (requiring at least 5 measurements at 1 GV, going up to 10 layers at 3 GV). The requirements result in a high selection efficiency with the low energy limit of the analysis of around 200 - 300 MeV/n.  For more details about different selection criteria and their resulting efficiencies see \cite{2016ApJ...818..1}. The selection efficiency of  this work will be discussed in section \ref{sec:isotopic_ratios}

In Fig.~\ref{im:calo_tm_r} the mean $dE/dx$ for each event vs. the rigidity measured with the magnetic spectrometer for $Z = 3$ and $Z = 4$ particles is shown. The energy loss in MeV was derived from the measurement in MIP using a conversion factor \citep{2016ApJ...818..1}. 
In Fig.~\ref{im:beta_r} and Fig.~\ref{im:calo_tm_r} the isotopic separation between \bese\ and (\beni+\bete) is clearly visible, while the separation between \lisi\ and \lise\ or \beni\ and \bete\ is difficult to see.

\subsubsection{\pam\ mass resolution} \label{sec:mass_resolution}

To separate isotopes, a good mass resolution is important: for two isotopes with a mass difference of one atomic mass unit (amu) a mass resolution of 0.15 amu (1-sigma) would allow a clear separation of two gaussian-like distributions without any overlap, a worse mass resolution results in an overlap of the two distributions, but two peaks would still be visible. When the resolution is worse than $\approx$ 0.4 amu, it is not possible to resolve the two peaks anymore. 

In a magnet spectrometer similar to \pam\ there are three independent contributions to the mass resolution : the bending power of the magnetic spectrometer coupled with the intrinsic spatial resolution limits of the tracking
detectors ($MDR_{spec}$), the multiple scattering of the particle along its path in the bending area of the magnet ($MDR_{coul}$), and the precision of the velocity measurement (given either by timing or by measuring the energy loss). For \pam\ the overall momentum resolution of the magnetic spectrometer  has been measured in beam tests at CERN for $Z = 1$ particles \citep {2007APh....27..296P}. From the measurements at high energies, where the contributions from multiple scattering is negligible, one can derive that the $MDR_{spec}$ of the \pam\ spectrometer has a value of about 1 TV. For $Z > 2$ particles the tracking algorithm was modified, which resulted in a more precise fitting and a higher track finding efficiency, but resulting in a worse spatial resolution \citep{2014ApJ...791..93}, which reduced the $MDR_{spec}$. An analysis of the residuals (difference between the fitted track and the measured positions) showed that the spatial resolution for lithium particles is comparable to the resolution for protons, while for beryllium particles the spatial resolution is about twice as large, resulting in an $MDR_{spec}$ of about 500 GV.

In both cases the contribution of $MDR_{spec}$ to the overall mass resolution is negligible for small rigidities (up to a few GV), here the contribution from multiple scattering is the dominant effect. Its value is inversely proportional to the bending power of the magnet ($\int B \cdot dl$) and direct proportional to the amount of matter traversed along the bending part of the track.
\pam\ combines a strong magnetic field of 0.43 T with a low amount of material (only the six silicon detectors, each 300 $\mu$m, total grammage 0.42 g/cm$^2$) in the magnetic cavity. This leads to a respective value for $MDR_{cou}$ around  3.5 \% at 8 GV, increasing to about 5\% at 1 GV for $Z = 1$ particles \citep{2007APh....27..296P}, which is in agreement with the \texttt{GEANT4} \pam\ simulation. For $Z = 3$ and $Z = 4$ particles, for which no test beam data were
available, the simulation provides values for $MDR_{cou}$ of  $\approx$ 3.8 \% at 5 GV, increasing to about 6 - 7\% at 1 GV. 

\begin{figure}[t]
    \centering
    \plottwo{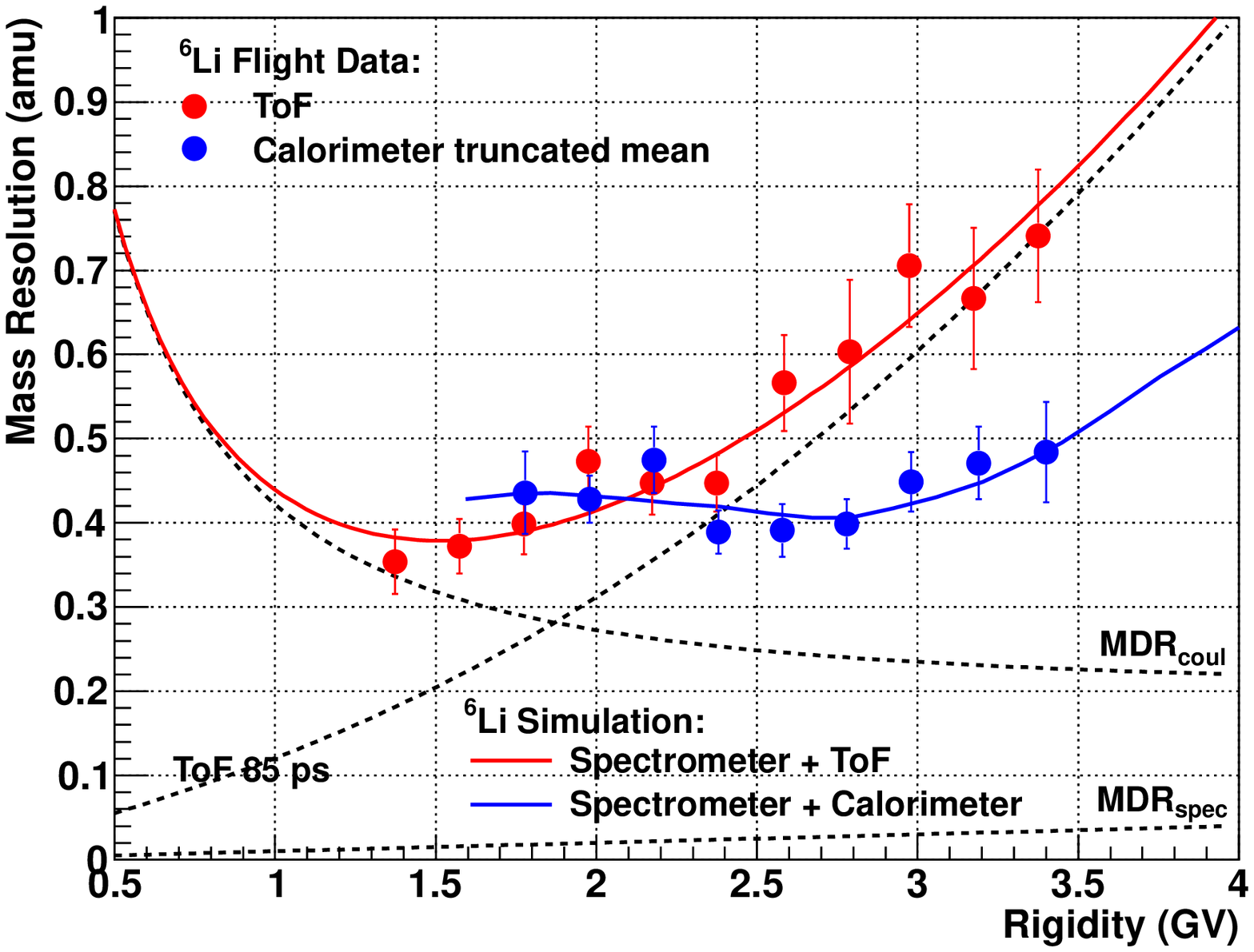}{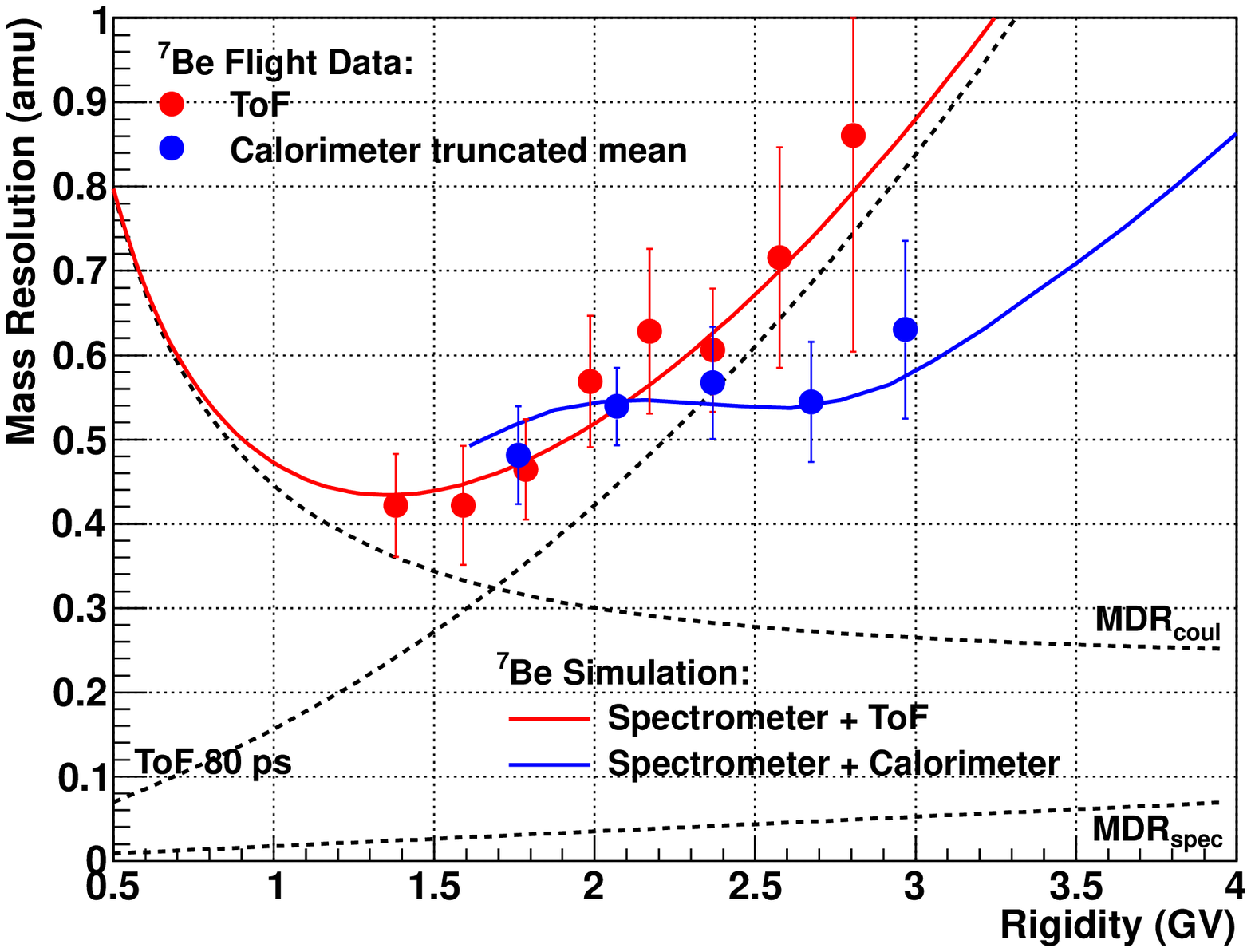}
    \caption{Measured \lisi\ (left) and \bese\ (right) mass resolution for the ToF (red circles) and the calorimeter using the ``truncated mean'' method (blue circles). The dashed lines show the calculated independent contributions (rigidity ($MDR_{spec}$), multiple scattering ($MDR_{cou}$), and velocity via ToF, while the solid red and blue lines show the overall mass resolution.}
\label{im:mass_resolution_li_be}
\end{figure}

In Fig.~\ref{im:mass_resolution_li_be} the three independent contributions to the mass resolution for \lisi\ and \bese\ particle from rigidity ($MDR_{spec}$), multiple scattering ($MDR_{cou}$), and velocity via ToF are shown as dotted lines, and the red curve showing the overall expected mass resolution of the $\beta_{ToF}$ versus rigidity technique. 
The blue curve displays the overall resolution for the multiple $dE/dx_{Calorimeter}$ versus rigidity technique  simulation (the independent contribution from the calorimeter alone is not shown).

To derive the mass resolution for flight data, a sample of \lisi\ or \bese\ particles was selected using strict cuts on the mass coming from the ToF or on the multiple $dE/dx$ from the calorimeter. Thus, the mass resolution of each detector was derived using the other one, assuming that the contamination from other isotopes is negligible. As it can be seen, the experimental results on the mass resolution follow the prediction from the \texttt{GEANT4} \pam\ simulation. Furthermore, the approach with multiple $dE/dx$ from the calorimeter extends the \pam\ measurements on isotopes to higher energies. It can also be noticed that the lower limit of the mass resolution in this analysis is determined by the multiple scattering.

\subsection{Raw isotope numbers with the ToF and calorimeter}\label{sec:raw_numbers}

The isotope separation was performed identical to \cite{2016ApJ...818..1} in intervals of kinetic energy per nucleon. Since the magnetic spectrometer measures the rigidity of particles, this implies different rigidity intervals according to mass of the isotope under study.

The expected distributions of the measured quantities ($dE/dx$ for the calorimeter and 1/$\beta$ for the ToF) in each rigidity interval were modeled and then likelihood fits were performed using the ``TFractionFitter'' method in \texttt{ROOT} \citep{root}. The \texttt{GEANT4} \pam\ simulation was used to derive the model distributions for the observables.

\begin{figure}[t]
    \centering
    \plottwo{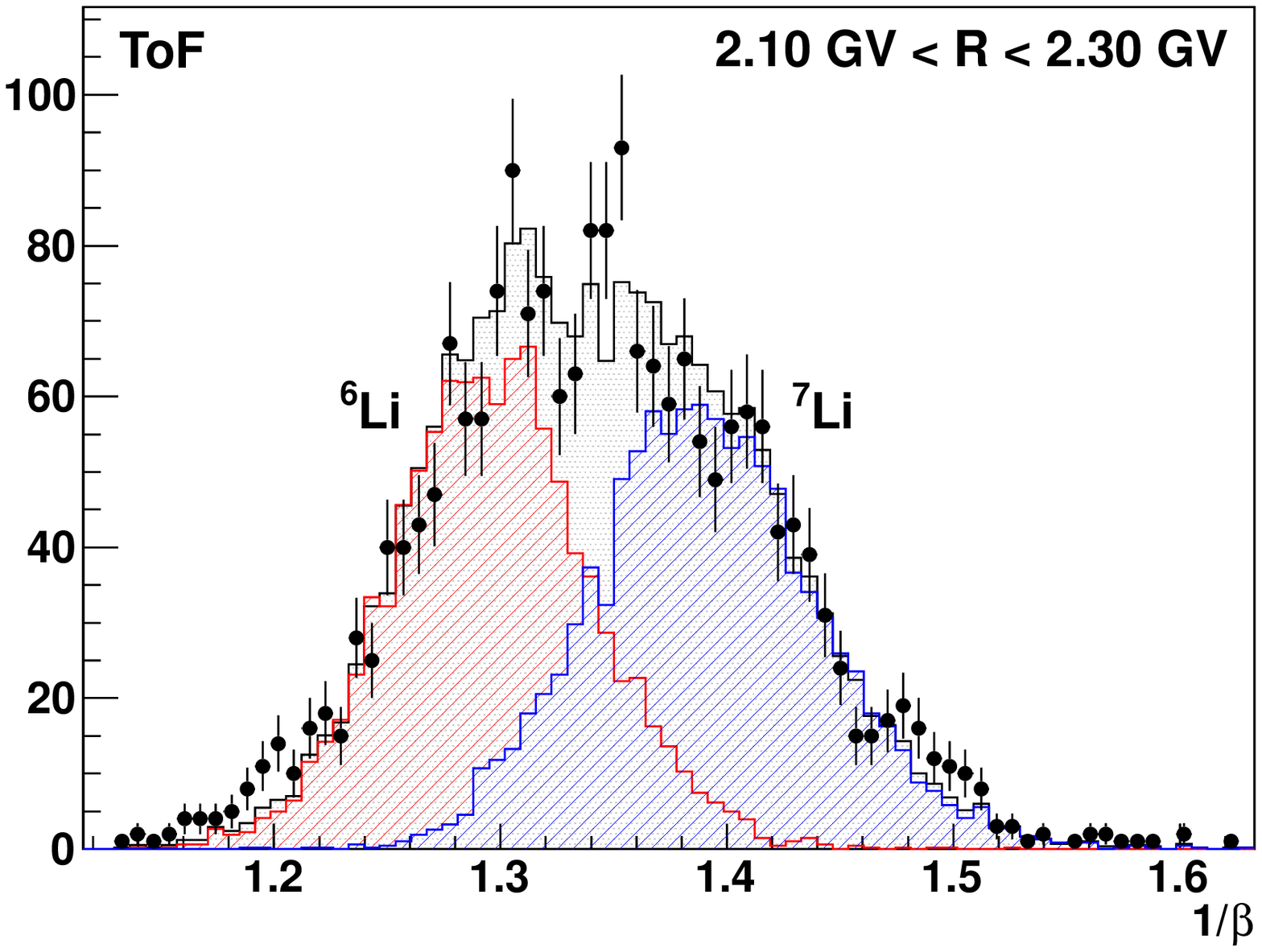}{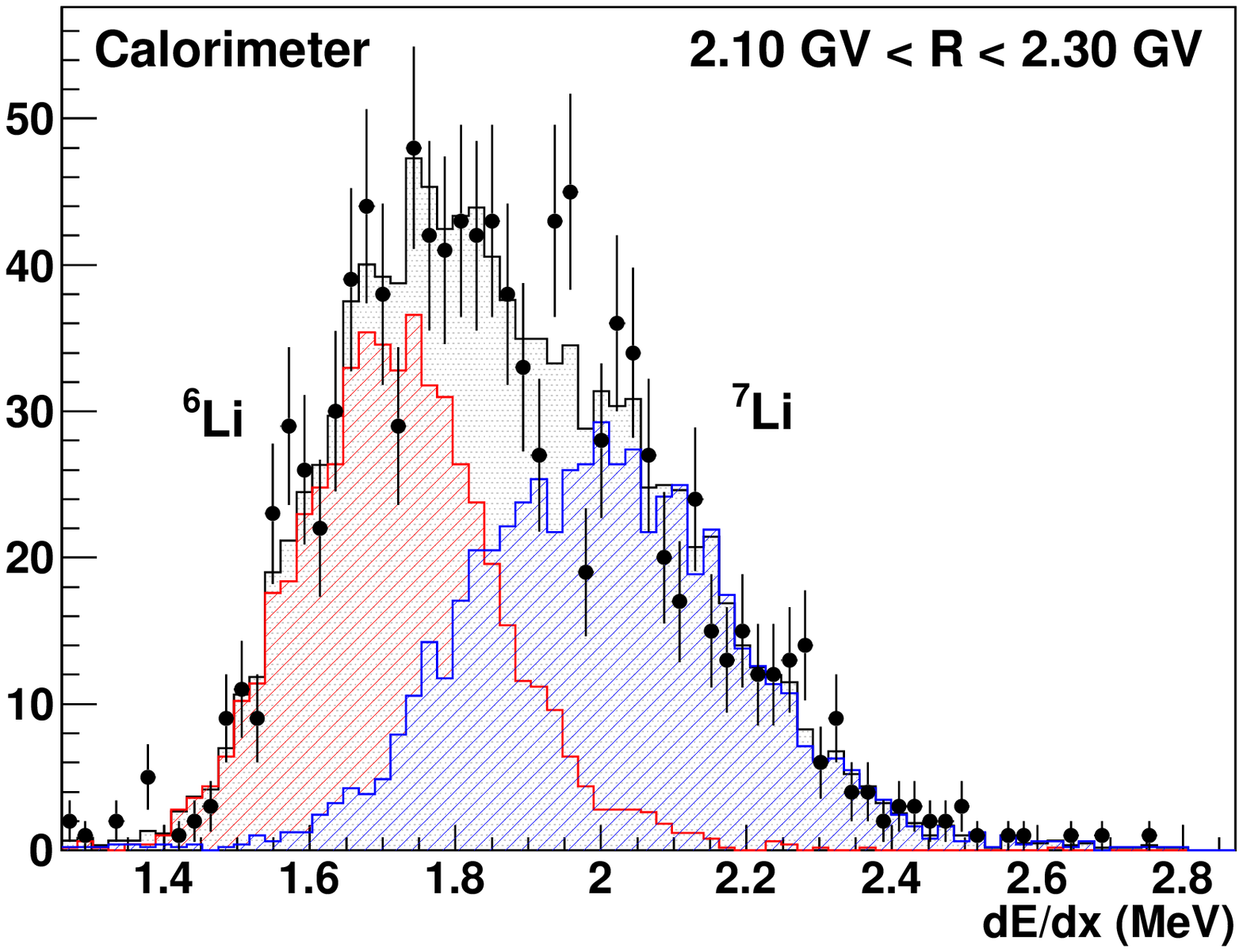}
    \caption{$1/\beta$ distributions of the ToF (left) and truncated mean $dE/dx$ of the calorimeter (right) for lithium in the 2.1 - 2.3 GV range.}
    \label{im:fitter_li}
\end{figure}

\begin{figure}[t]
    \centering
    \plottwo{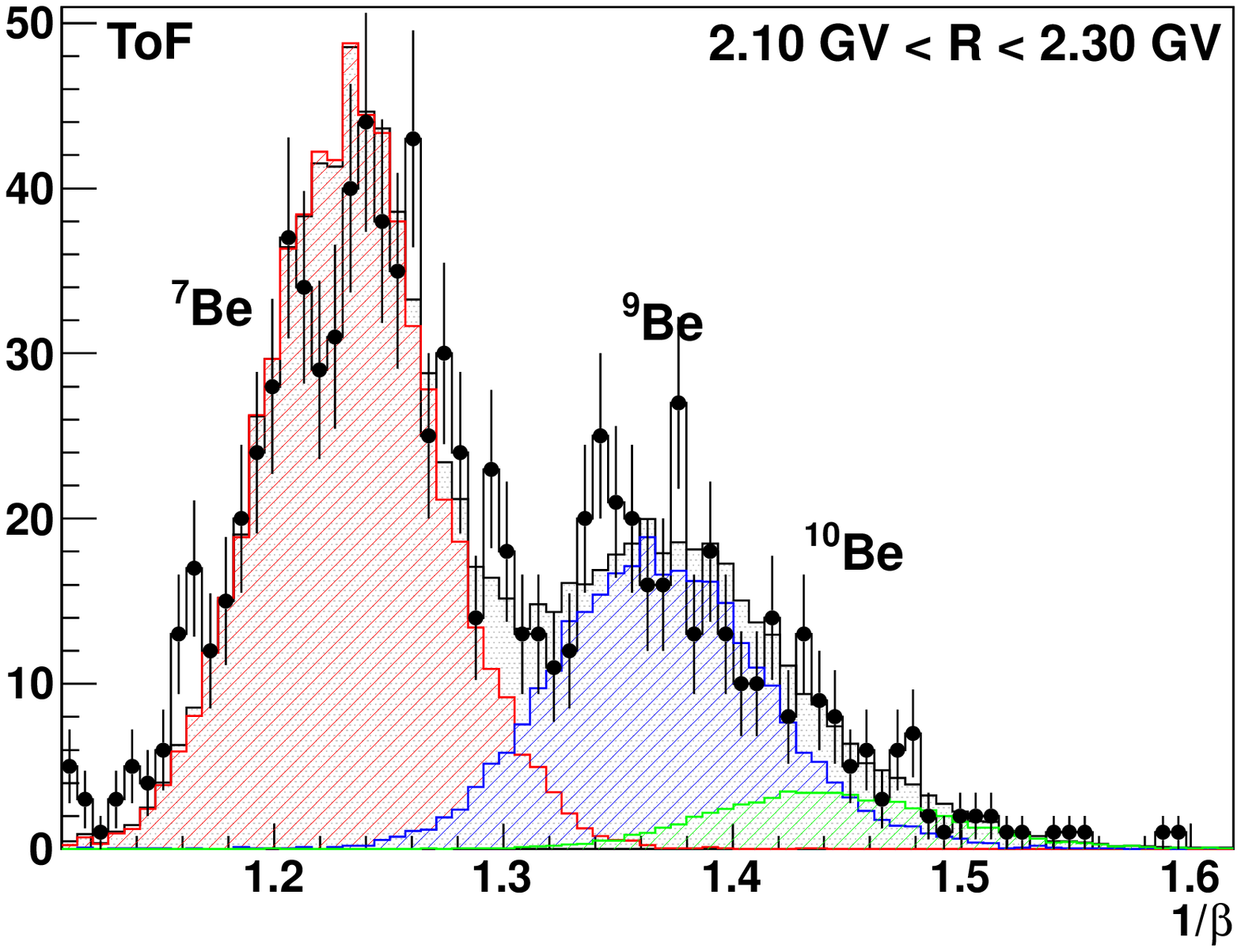}{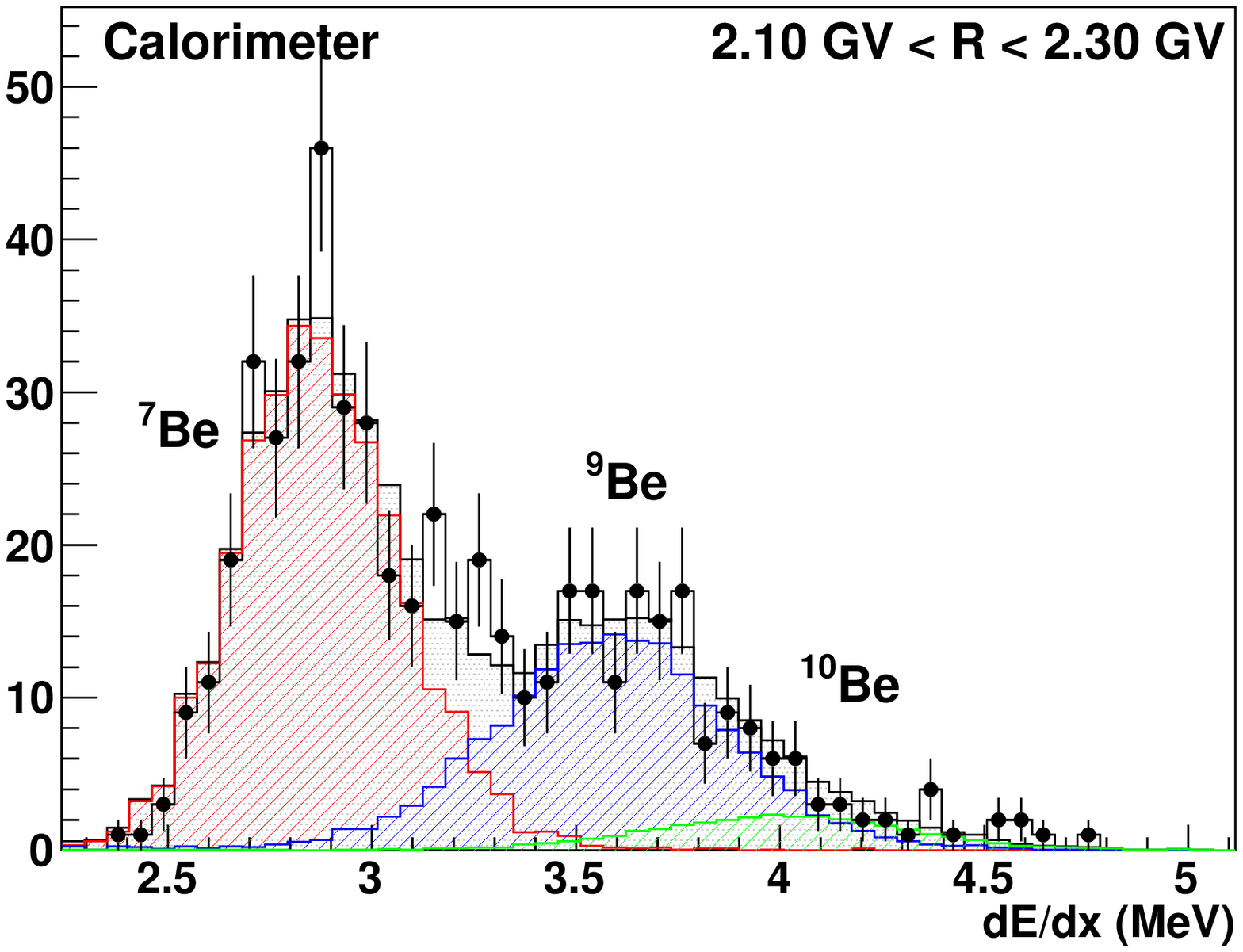}
    \caption{$1/\beta$ distributions of the ToF (left) and truncated mean $dE/dx$ of the calorimeter (right) for beryllium in the 2.1 - 2.3 GV range.}
    \label{im:fitter_be}
\end{figure}

As an example in Fig.~\ref{im:fitter_li} the distributions of ToF and calorimeter for lithium in the 1.9 - 2.1 GV rigidity range are shown, and in Fig.~\ref{im:fitter_be} similar distributions for beryllium. The grey area shows how the combined fit using the two (or  three) model distributions derived with the modified \texttt{GEANT4} \pam\ simulation matches the data points (black points) while the colored areas shows the estimated individual isotope signals.

\subsubsection{Tuning of the simulation}\label{sec:tune_simulation}

When analyzing hydrogen and helium isotopes with the calorimeter, taking the simulated energy loss in each layer as coming from the \texttt{GEANT4} \pam\ simulation, it was noticed that the resulting distributions showed a slight mismatch with the flight data. It was found that the widths of the histograms were smaller than in the real data, also there was a small offset. A multiplicative factor to the simulated energy loss in a layer was applied, plus adding a gaussian spread of the signal of a few percent \citep{2016ApJ...818..1}. These corrections were used also for the analysis of lithium and beryllium.
Small deviations were also found for the 1/$\beta$ distributions of the ToF, which is understandable, since the creation of the timing signals in our simulation is somewhat simplified (for example no light tracing in the scintillators) compared to the real physical processes.

As one can see from Fig~\ref{im:mass_resolution_li_be}, the mass resolution of the simulated data is in good agreement with the flight data, thus the main task was to find the offset between simulated and flight data distributions.

\begin{figure}[t]
    \centering
    \plottwo{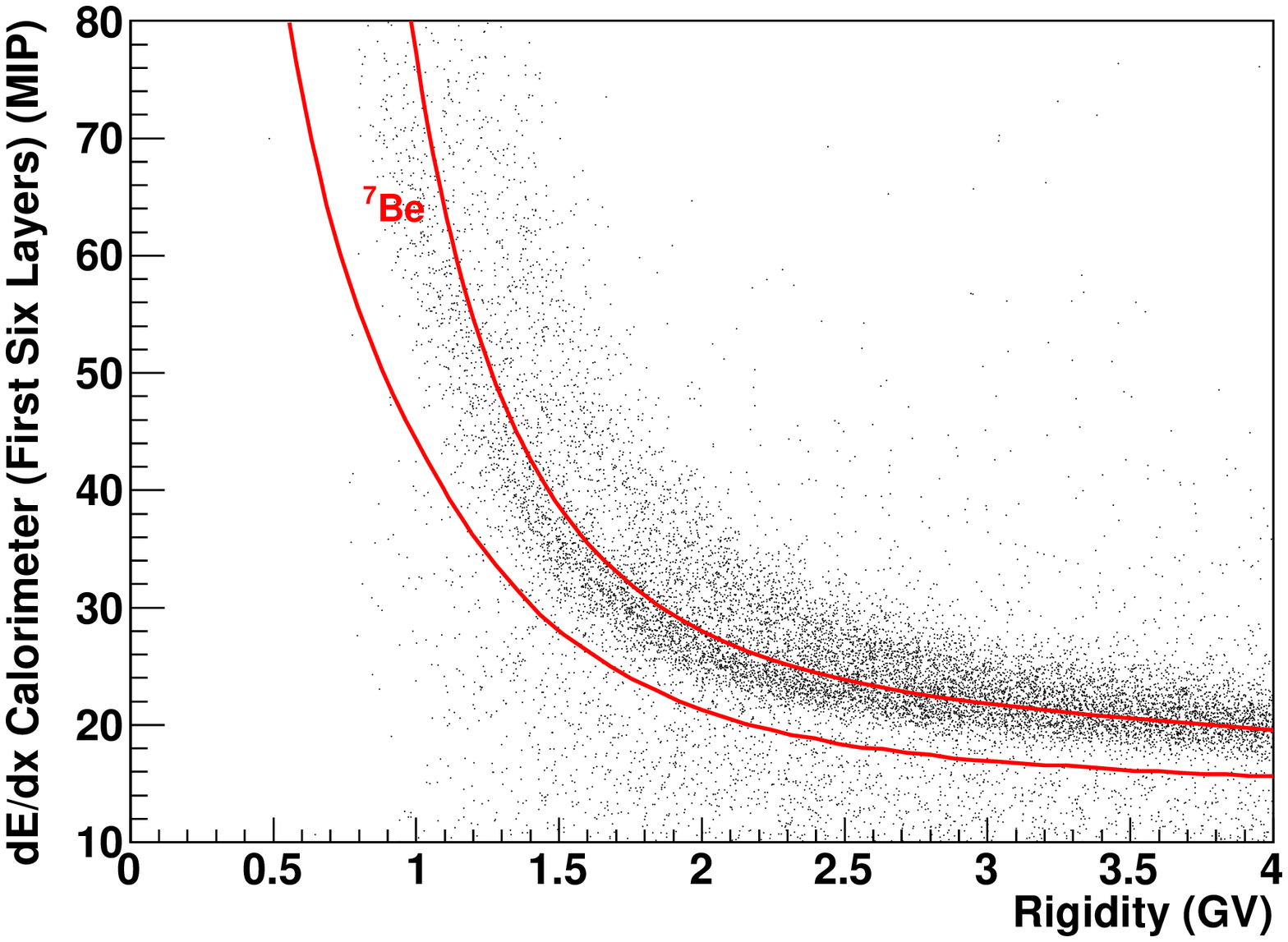}{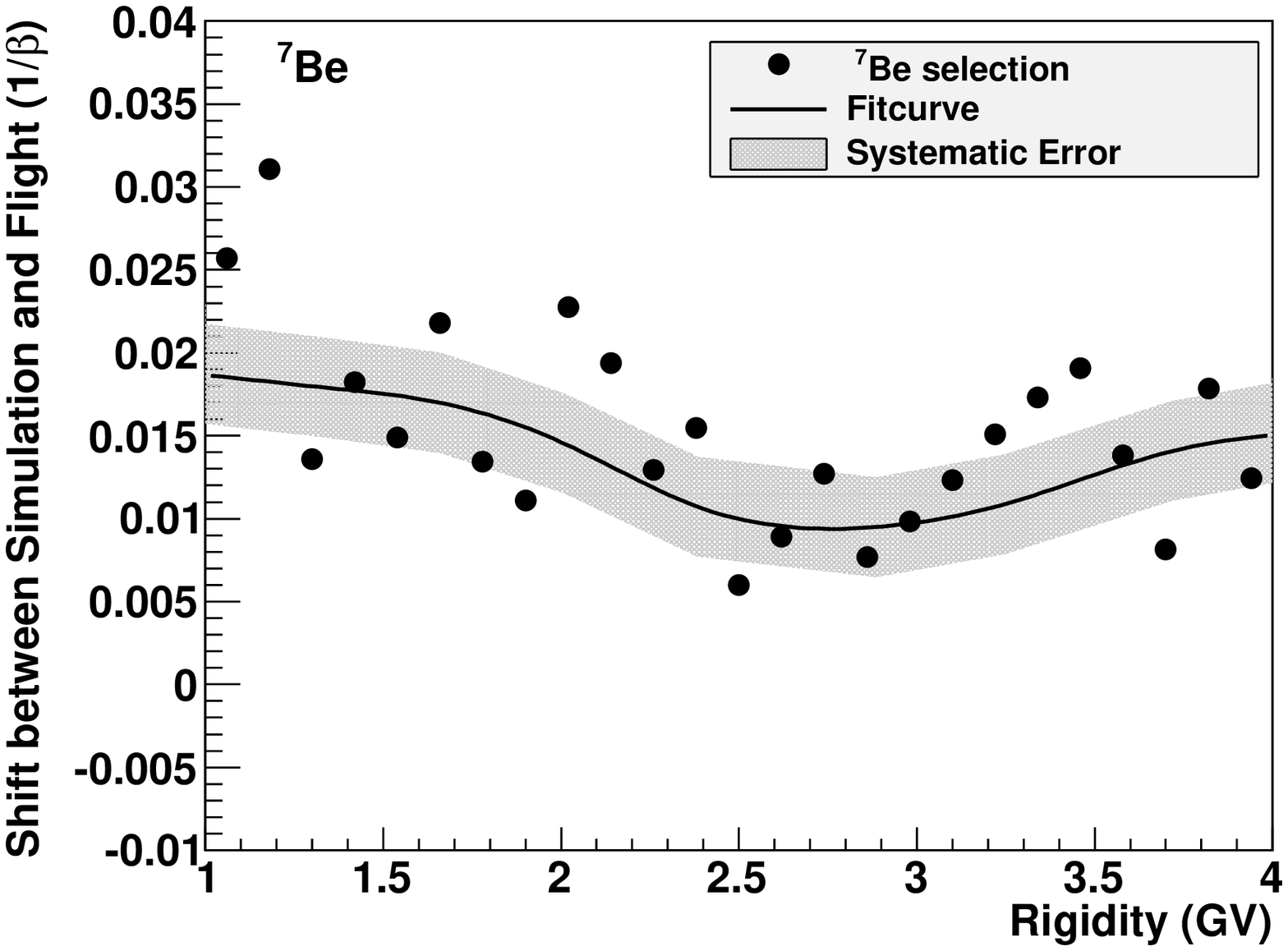}
\caption{Left: Selection of a \bese\ sample in the flight data using the truncated mean of the calorimeter (only first six layers used), the selection is depicted by the red solid lines. Right: ToF 1/$\beta$ ``shift'' for  \bese\ (black circles) together with a best-fit curve and the estimated systematic error band}
    \label{im:tof_shift_be7}
\end{figure}

Starting with the beryllium analysis, redundant detectors were used to separate \bese\ samples from the flight data. As an example the selection with the calorimeter is shown in Fig.~\ref{im:tof_shift_be7} on the left, with the actual selection of \bese\ particles depicted by the red solid lines. To make the selection of low rigidity events possible, a special truncated mean value of $dE/dx$ was used, taking only the first six layers of the calorimeter into account. It is obvious that the \bese\ sample will be affected by some contamination from \beni\ and \bete, especially at higher rigidities.
Thus the full simulated data set was created by summing up events from the single \bese, \beni, and \bete\ datasets, with the number of events weighted according to their expected abundance. After applying the same \bese\ calorimeter selection cuts to the flight data and the simulated data, the 1/$\beta$ distributions of the selected \bese\ samples were compared in rigidity intervals. The difference of the two mean 1/$\beta$ values (``shift'') in each rigidity interval is shown in Fig.~\ref{im:tof_shift_be7} on the right (black dots) together with a best-fit function. The deviation of the fitpoints from the best-fit curve in Fig.~\ref{im:tof_shift_be7} is used to estimate the systematic error of the shift function (RMS of the spread distribution of the points from the curve).

The shift curve derived for \bese\ was then finally used to correct the 1/$\beta$ values for all the simulated beryllium events, since it is not possible to get the functions for \beni\ and \bete\ out of the data. This might introduce some systematic error to the calculation of the individual \beni\ and \bete\ numbers.
In a similar manner the shift function for the truncated mean of the calorimeter was derived, using the ToF to select the \bese\ sample.

The shift functions for the lithium analysis were derived in the same way as for beryllium, this time using the redundant detectors to select a \lisi\ and \lise\ sample. Since \lisi\ and \lise\ are neighbor isotopes, great care was taken to minimize the effects of the selection and the contamination of one isotope into the other.
Finally the shifts for \lisi\ and \lise\ were not used individually, but only one best-fit curve was derived for the further analysis, the differences were treated as systematic errors. 

As already written, the shift functions were finally used to correct the simulated data event by event, resulting in the distributions shown in Fig. \ref{im:fitter_li} and Fig. \ref{im:fitter_be}.

\subsection{Determination of isotopic ratios}\label{sec:isotopic_ratios}
 
To derive the isotopic ratios, the numbers of selected events derived with the ``TFractionFitter'' likelihood procedure had to be corrected for the selections efficiencies and particle losses. 

\begin{figure}[t]
\begin{minipage}[t]{0.33\textwidth}
\includegraphics[width=\textwidth]{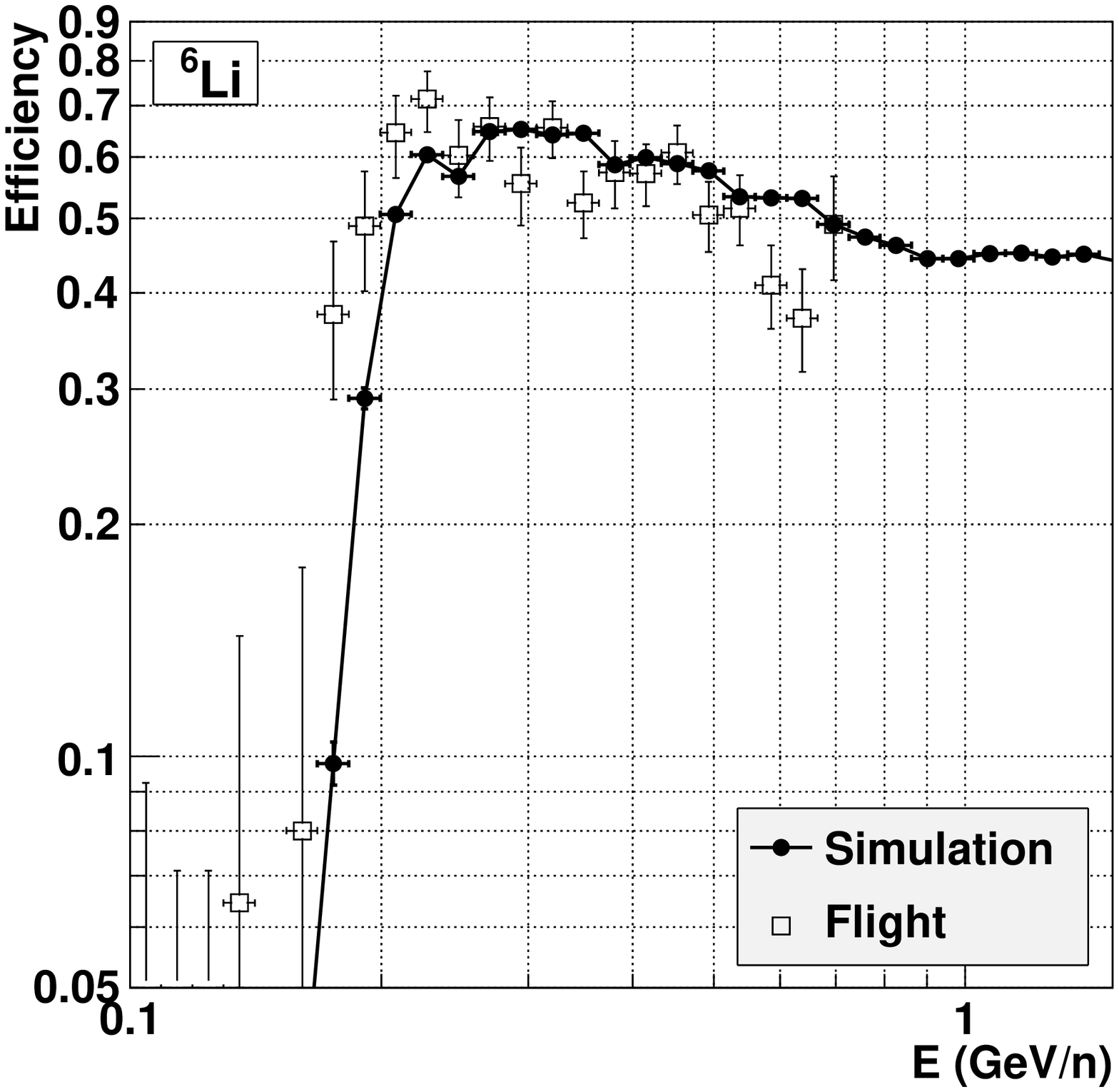}
\end{minipage}
\begin{minipage}[t]{0.33\textwidth}
\includegraphics[width=\textwidth]{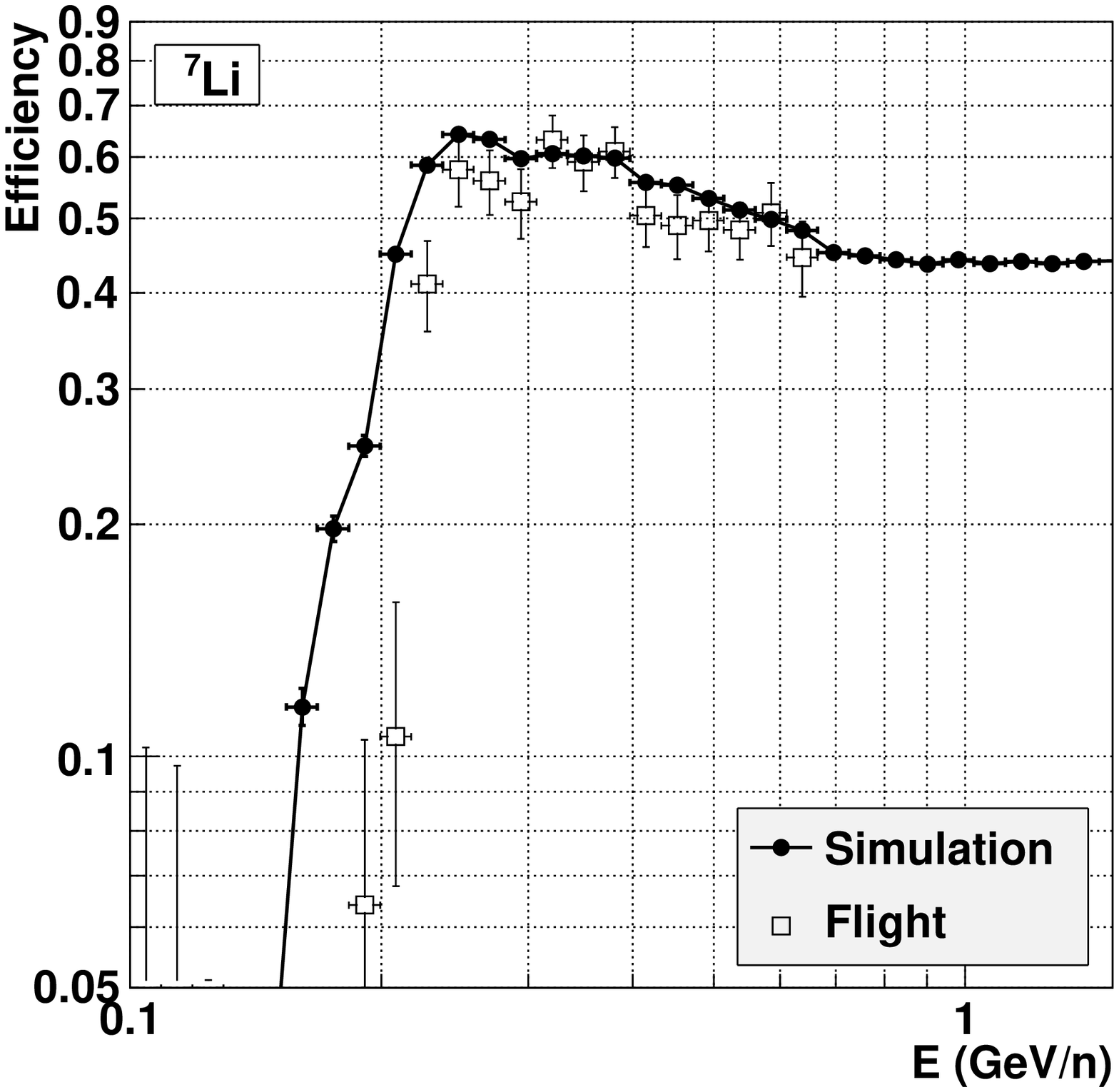}
\end{minipage}
\begin{minipage}[t]{0.33\textwidth}
\includegraphics[width=\textwidth]{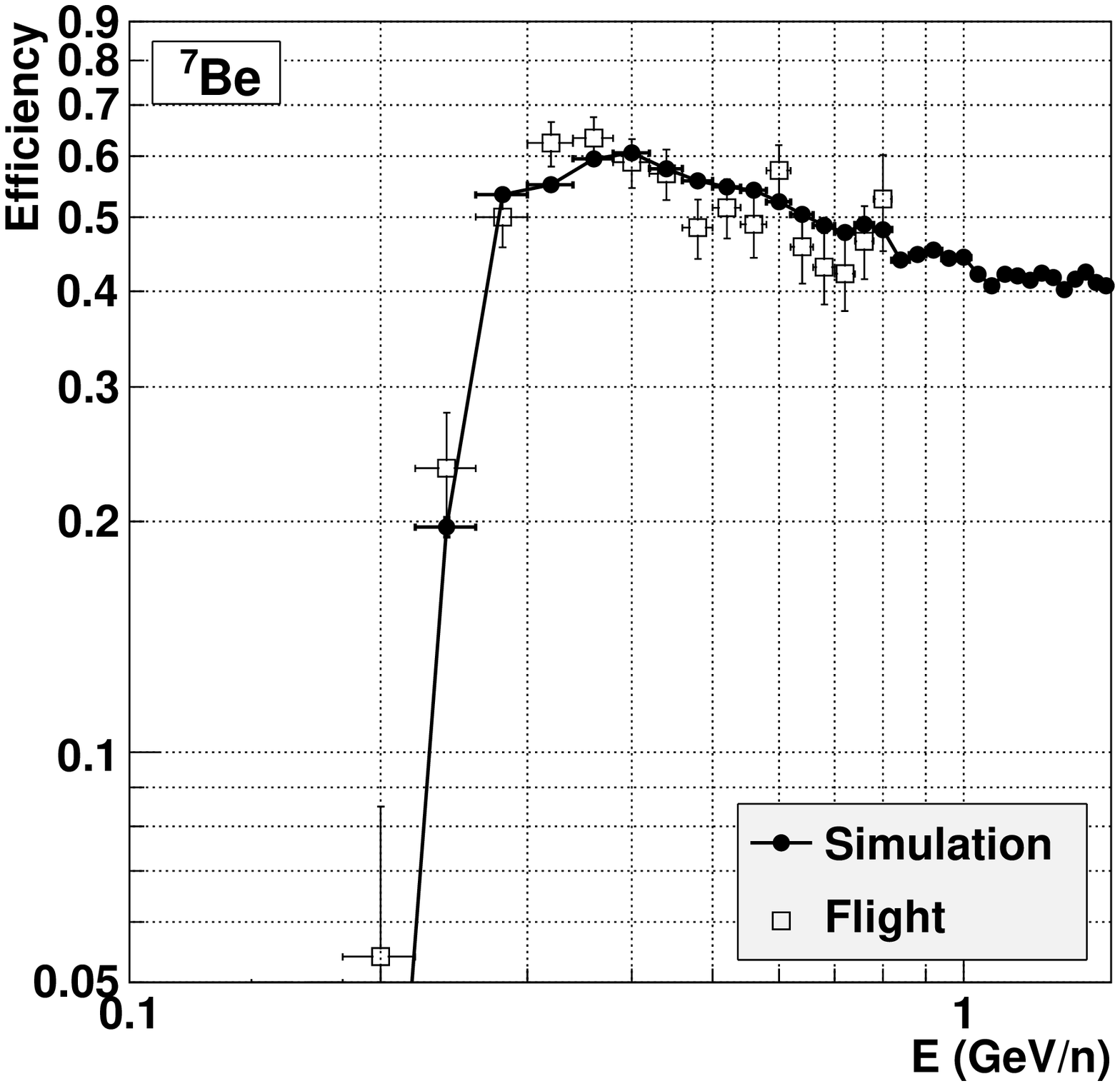}
\end{minipage}
\caption{Comparison between the calorimeter efficiency derived with simulated data (circles) and flight data (squares) for \lisi\ (left), \lise\ (middle), and \bese\ (right)}
\label{im:calo_effi}
\end{figure}

The efficiencies were mostly derived using simulations, these results were then checked with flight data by using redundant detectors to create test samples. For example the efficiency of the charge selections using S12, $\langle$S2$\rangle$, and $\langle$S3$\rangle$ was evaluated on a sample of events selected using the charge information of S11 and the calorimeter, similar to the method used for the analysis of boron and carbon fluxes \citep{2014ApJ...791..93}.
The efficiency of the tracking system was however obtained from the Monte Carlo simulation.
The comparison between calorimeter efficiencies derived with simulated data and the ones derived with flight data (using a selection combining the $\beta$ and $dE/dx$ information from the ToF with the rigidity from the magnetic spectrometer) is shown in Fig.~\ref{im:calo_effi}.

The calorimeter selection used for this analysis results in a quite flat efficiency of about 50 - 60 \% above 200 - 300 MeV/n, for lower energies the effiency shows a steep drop, since the particles stop already in the top layers of the calorimeter. 
As one can see in Fig.~\ref{im:calo_effi}, there is a good agreement between the simulation and flight data for the higher energies. At low energies the agreement is very good for \bese\, while for \lisi\ and \lise\ there are some differences to see. 
A similar behaviour was found in \cite{2016ApJ...818..1} for hydrogen and helium isotopes. It seems that at lower energies, where our selection cuts for the truncated mean are quite soft and only a small number of layers are used to derive the truncated mean, fluctuations are probably still significant, and the \texttt{GEANT4} \pam\ simulation cannot fully reproduce the actual energy loss under these circumstances.
However, the calorimeter selection was used only above 300 MeV/n. For the calculation of the isotopic ratios the simulated efficiency was used.

Furthermore one needs the geometrical factor and the live time of the instrument as evaluated by the trigger system \citep{2013ApJ...770..2}, since these values depend on the rigidity.
The nominal geometrical factor $G_F$ of \pam\ is almost constant above 1 GV, with the requirements on the fiducial volume corresponding to a value of $G_F = 19.9$  cm\textsuperscript{2} sr.
The live time depends on the orbital selection as described in section \ref{sec:event_selection} and is evaluated by the trigger system \citep{bru08}. 

The finite resolution of the magnetic spectrometer and particle slowdown due to ionization energy losses results in a distortion of the particle spectra, which affects the isotopic ratio. We used a Bayesian unfolding procedure 
\citep{dagostini} to correct for this effect (see \cite{2011Sci...332...69A}).

The particle slowdown in the instrument leads also to a loss of galactic particles, since they do not pass the selection cut in rigidity ($\rho$ $>$ 1.3 $\rho_{SVC}$). At 150 MeV/n the loss is about 10\%, decreasing to just 1\% already at 450 MeV/n. This effect is accounted for using the Monte Carlo simulations.

Because of hadronic interactions in the instrument lithium and beryllium nuclei might be lost. The energy dependent correction factor has been derived using the \texttt{GEANT4} \pam\ simulation. The correction factors at 1 GeV/n are $\simeq 17\%$ for lithium and $\simeq 19\%$ for beryllium, with small differences (roughly 1 - 3\%) between the isotopes. 

The selected lithium and beryllium samples are contaminated by secondaries, which are produced in fragmentation processes occurring in the aluminum dome on top of the pressurized vessel. 
As done in the analysis of the boron and carbon spectra \citep{2014ApJ...791..93} the contamination was studied with a Monte Carlo calculation based on the FLUKA code \citep{battistoni}. Cosmic spectra for oxygen, carbon, boron, and beryllium were simulated, then scaled to the expected numbers at the top of the instrument. We found that roughly 1.5 times more \lise\ than \lisi\ is produced, but since the number of produced \lisi\ or \lise\ is less than 1\% of the selected flight data sample, the effect on the \lise/\lisi\ ratio is negligible.
The fragmentation of oxygen and carbon into beryllium creates roughly the same numbers of the three beryllium isotopes, however the fragmentation of {\textsuperscript{10}B} and {\textsuperscript{11}B} creates significantly more \beni\ and \bete\ than \bese.  After subtracting the contamination, the \bese/(\beni\ + \bete) ratio increases by $\simeq$ 7\% at 200 MeV/n, while the difference is $\simeq$ 3\% at 1 GeV/n. 

\subsection{Systematic uncertainties}\label{sec:systematics}

Systematic uncertainties were estimated as discussed in \citep{2016ApJ...818..1}.
The event selection criteria described in section \ref{sec:event_selection} are less strict compared to our previous works \citep{2011Sci...332...69A, 2016ApJ...818..1}, where we quoted a systematic uncertainty of 3.6\%.
However, it was found in \cite{2013ApJ...765..91} that less stringent cuts implied a larger systematic error. For this work a systematic error of 5\% was derived.

The correction for particles lost due to the selection of galactic particles (see section \ref{sec:event_selection}) has an uncertainty due to the size of the Monte Carlo sample. The systematic error decreases from $3\%$ at 100 MeV/n to $0.05\%$ at 1000 MeV/n.

Using the \texttt{GEANT4} \pam\ simulation it was studied how the uncertainties on the modeled distribution affects the reconstructed number of events. It was found that the dominant effect was the shifts of the modeled distributions. For example, the systematic uncertainty of the shift for the truncated mean of the calorimeter transforms in a systematic error of the measured \lisi\ or \lise\ numbers of roughly 20\%.

The efficiency of the calorimeter selection was derived using simulated and flight data, as shown in Fig.~\ref{im:calo_effi}. The agreement between the two methods is quite good, and we assigned a conservative systematic error of $3\%$ independent from the energy.

The systematic uncertainties of the unfolding procedure (2\%, independent of energy) and the effective geometrical factor ($0.18\%$, practically independent of energy) are taken from \cite{2016ApJ...818..1} without changes and we refer to this paper for more details.

The uncertainty of the correction for particles lost due inelastic interactions is $\simeq$ $1\%$ due to the size of the Monte Carlo sample.

\section{Results and discussion}

\begin{figure}[t]
    \centering
    \plotone{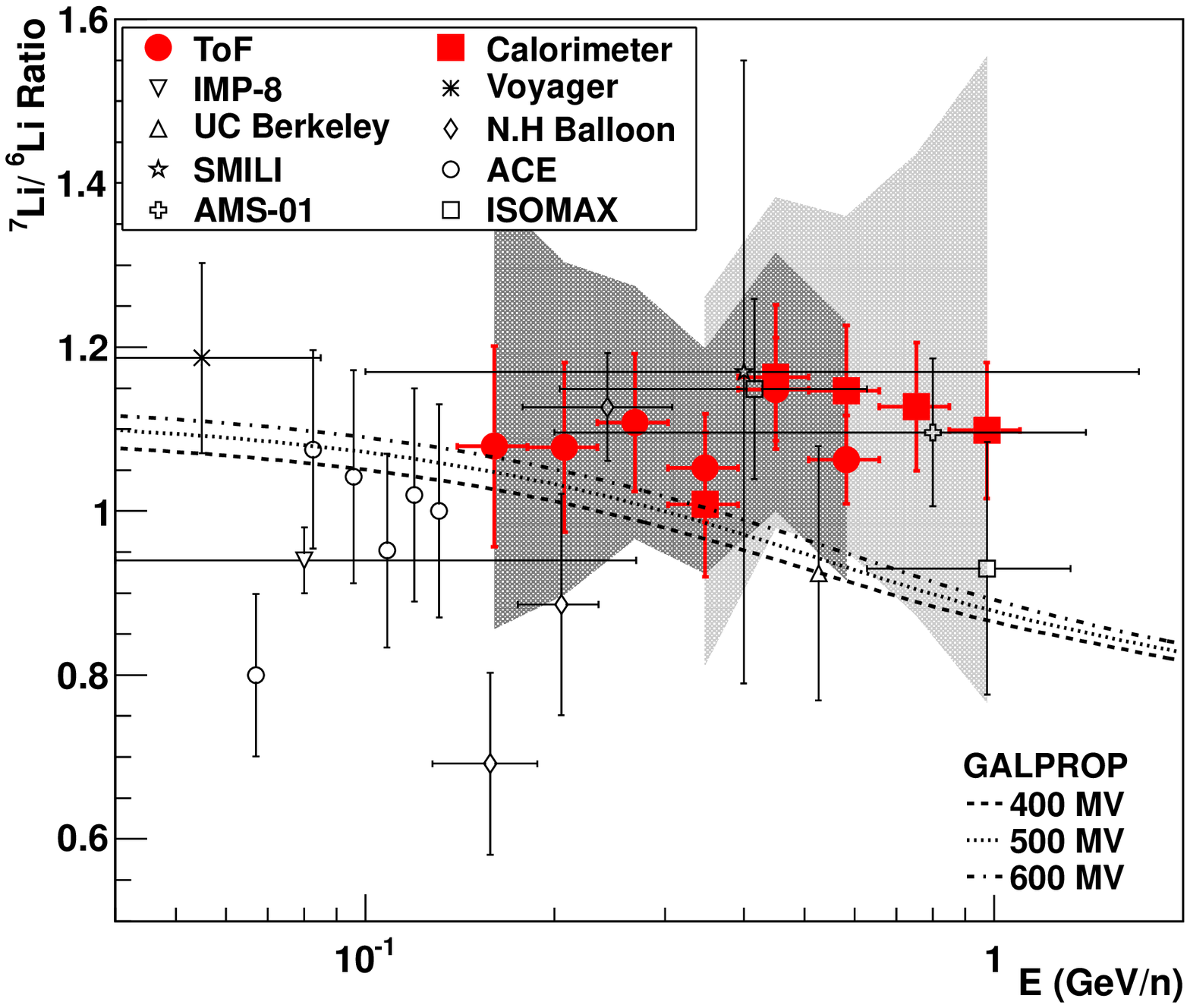}
    \caption{Ratio of \lise/\lisi\ derived with the \pam\ ToF (circles) or the calorimeter (squares). Error bars show the statistical uncertainty while shaded areas show the systematic uncertainty.
		Previous experiments are: 
		IMP-8 \citep{1975ApJ...201..L145}, UC Berkeley \citep{1978ApJ...226..355B}, NH-Balloon \citep{NH-balloon},
		Voyager \citep{2002ApJ...568..210W}, SMILI \citep{2000ApJ...534..757A}, 
		ACE \citep{ACE2006},		
		 AMS-01 \citep{2011ApJ...736..105A}, ISOMAX \citep{2004ApJ...611..892H}.
		Also shown are predictions of GALPROP webRun v54.1 \citep{2011CoPhC.182.1156V} using di†ferent
   solar modulation parameters. }
    \label{im:li_results}
\end{figure}

\begin{figure}[t]
    \centering
    \plotone{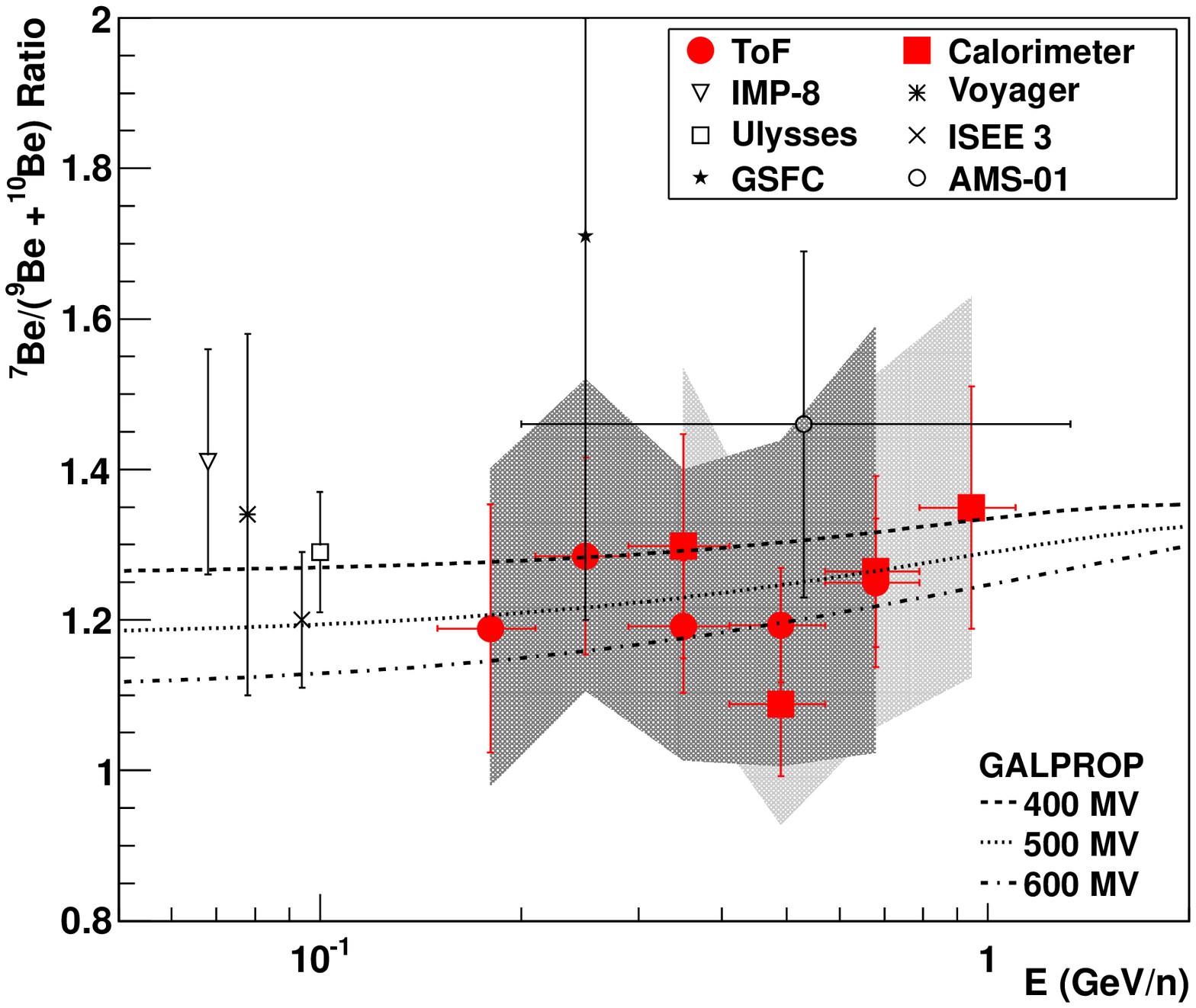}
		\caption{Ratio of \bese/(\beni\ + \bete) derived with the \pam\ ToF (circles) or the calorimeter (squares). Error bars show the statistical uncertainty while shaded areas show the systematic uncertainty.
Previous experiments are: Voyager \citep{2002ApJ...568..210W}, 
ULYSSES \citep{1998ApJ...501L..59C}, ISEE3 \citep{1980ApJ...239L.139W}, IMP7/8 \citep{1977ApJ...217..859G}, GSFC \citep{1977ApJ...212..262H}, AMS-01 \citep{2011ApJ...736..105A}. 
Also shown are predictions of GALPROP webRun v54.1 \citep{2011CoPhC.182.1156V} using di†ferent
solar modulation parameters. }
    \label{im:be_results}
\end{figure}

In the following the results for isotope ratios lithium and beryllium together with other measurements are shown, in Fig.~\ref{im:li_results} the \lise/\lisi\ ratio and in Fig.~\ref{im:be_results} the \bese/(\beni+\bete) ratio.  Results are also reported in Tables 
\ref{tab:lithium_table} and \ref{tab:beryllium_table}. 
As described in \ref{sec:isotopic_ratios}, the calorimeter efficiency decreases steeply below 200 - 300 MeV/n, thus we do not provide results of the calorimeter method for the lowest energy bins. Similarly for the ToF at the highest energy bins: here the mass resolution of the ToF is significantly worse compared to the calorimeter (see \ref{sec:mass_resolution}). 
The \pam\ results obtained via the ToF analysis and via the multiple $dE/dx$ measurements with the calorimeter agree very well within their systematic errors, giving confidence to the results.

\renewcommand{\arraystretch}{1.5}

\begin{deluxetable*}{ccc}
\tablecaption{$^7$Li/$^6$Li ratio derived with the \pam\ ToF and calorimeter, errors are statistical and systematics respectively. \label{tab:lithium_table}}
\tabletypesize{\scriptsize}
\tablewidth{0pt}
\tablehead{
\multicolumn{1}{c}{Kinetic energy} & \multicolumn{1}{c}{$^7$Li/$^6$Li ToF} & \multicolumn{1}{c}{$^7$Li/$^6$Li Calorimeter} \\
\multicolumn{1}{c}{at top of payload} & \multicolumn{1}{c}{value $\pm$ stat. $\pm$ syst.} & \multicolumn{1}{c}{value$\pm$ stat. $\pm$ syst.}\\
\multicolumn{1}{c}{(GeV n$^{-1}$)} & \multicolumn{2}{c}{}\\
}
\startdata
0.140 - 0.181 & $(1.08 \pm 0.12 \, \substack{+{0.31} \\ -{0.22}})$ & - \\ 
0.181 - 0.234 & $(1.08 \pm 0.10 \, \substack{+{0.23} \\ -{0.18}})$ & - \\ 
0.234 - 0.303 & $(1.11 \pm 0.08 \, \substack{+{0.17} \\ -{0.14}})$ & - \\ 
0.303 - 0.392 & $(1.05 \pm 0.07 \, \substack{+{0.15} \\ -{0.13}})$ & $(1.01 \pm 0.09 \, \substack{+{0.25} \\ -{0.19}})$ \\ 
0.392 - 0.508 & $(1.15 \pm 0.06 \, \substack{+{0.17} \\ -{0.15}})$ & $(1.16 \pm 0.09 \, \substack{+{0.22} \\ -{0.18}})$ \\ 
0.508 - 0.657 & $(1.06 \pm 0.05 \, \substack{+{0.17} \\ -{0.15}})$ & $(1.15 \pm 0.08 \, \substack{+{0.21} \\ -{0.19}})$ \\ 
0.657 - 0.850 & - & $(1.13 \pm 0.08 \, \substack{+{0.31} \\ -{0.25}})$ \\ 
0.850 - 1.100 & - & $(1.10 \pm 0.08 \, \substack{+{0.46} \\ -{0.33}})$ \\ 
\enddata
\end{deluxetable*}

\begin{deluxetable*}{ccc}
\tablecaption{$^7$Be/($^9$Be + $^{10}$Be) ratio derived with the \pam\ ToF and calorimeter, errors are statistical and systematics respectively. \label{tab:beryllium_table}}
\tabletypesize{\scriptsize}
\tablewidth{0pt}
\tablehead{
\multicolumn{1}{c}{Kinetic energy} & \multicolumn{1}{c}{$^7$Be/($^9$Be+ $^{10}$Be) ToF} & \multicolumn{1}{c}{$^7$Be/($^9$Be+ $^{10}$Be) Calorimeter} \\
\multicolumn{1}{c}{at top of payload} & \multicolumn{1}{c}{value $\pm$ stat. $\pm$ syst.} & \multicolumn{1}{c}{value $\pm$ stat. $\pm$ syst.}\\
\multicolumn{1}{c}{(GeV n$^{-1}$)} & \multicolumn{2}{c}{}\\
}
\startdata
0.150 - 0.210 & $(1.19 \pm 0.16 \, \substack{+{0.21} \\ -{0.21}})$ & - \\ 
0.210 - 0.290 & $(1.28 \pm 0.13 \, \substack{+{0.23} \\ -{0.18}})$ & - \\ 
0.290 - 0.410 & $(1.19 \pm 0.09 \, \substack{+{0.21} \\ -{0.18}})$ & $(1.30 \pm 0.15 \, \substack{+{0.24} \\ -{0.19}})$ \\ 
0.410 - 0.570 & $(1.19 \pm 0.08 \, \substack{+{0.25} \\ -{0.19}})$ & $(1.09 \pm 0.10 \, \substack{+{0.19} \\ -{0.16}})$ \\ 
0.570 - 0.790 & $(1.25 \pm 0.09 \, \substack{+{0.34} \\ -{0.23}})$ & $(1.26 \pm 0.13 \, \substack{+{0.26} \\ -{0.21}})$ \\ 
0.790 - 1.100 & - & $(1.35 \pm 0.16 \, \substack{+{0.28} \\ -{0.22}})$ \\ 
\enddata
\end{deluxetable*}

In Fig.~\ref{im:li_results}  and Fig.~\ref{im:be_results} also results of the propagation model GALPROP (diffusion-halo model with reacceleration) as described in detail elsewhere \citep{1998ApJ...509..212S} are shown.
In this work the GALPROP webRun (https://galprop.stanford.edu/webrun) v54.1 version \citep{2011CoPhC.182.1156V} with the default input parameters was used.

The isotopic ratios are effected by solar modulation: in the years 2006 - 2014 the solar modulation parameter $\Phi$ varies between \~ 370 and \~ 750 MV \citep{Koldobskiy2018}. Since the efficiency of the tracking sytem decreased throughout the years, the majority of the flight data used for this analysis was collected in the years 2006 - 2009 at solar minimum conditions. Weighting the annual \pam\ count rate with the annual values of the solar modulation parameter $\Phi$ from \cite{Koldobskiy2018}, a mean value of 493 MV for $\Phi$ was calculated for this analysis. Three different curves of the GALPROP results for solar modulation parameters of 400, 500, and 600 MV are shown in Fig.~\ref{im:li_results} and Fig.~\ref{im:be_results}, as it can be seen the changes in the resulting ratios are relatively small compared to the other uncertainties.

The \lise/\lisi\ ratio measured by \pam\ matches the GALPROP prediction quite well at lower energies, while for higher energies the measured ratio is about 20\% higher than predicted. However, also the measured ratio of SMILI \citep{2000ApJ...534..757A}, AMS-01 \citep{2011ApJ...736..105A} and ISOMAX \citep{2004ApJ...611..892H} are somewhat higher than the prediction.
The model uncertainties are mostly due to uncertainties in the cross-sections of tertiary interactions such as the production of beryllium and lithium from isotopes of lithium, beryllium, and boron. Some important cross sections are poorly measured or even absent, see \citep{2003ICRC....4.1969M} for more details.

As is was already written, a clean selection of \beni\ and \bete\ is very difficult due to the mass resolution and low statistics, the results for the \bete/\beni\ ratio would suffer from a big systematic error. Thus we present only the \bese/(\beni+\bete) ratio.

It is interesting to point out that this ratio is in good agreement with the model predictions, while the results for \lise/\lisi\ (especially at higher energies) appear systematically higher.
This discrepancy implies to some extent that indeed specific cross sections or the model need to be investigated. However, a more comprehensive and detailed interpretation of our results in this context is beyond the scope of this paper.

\section{Acknowledgement}
We acknowledge support from the Russian Space Agency (Roscosmos),  the Russian Ministry of Education and Science, projects \No 3.2131.2017 and \No 1.5103.2017/\foreignlanguage{russian}{ВУ}, the Russian Foundation for Basic Research \No 16.02.00093 and \No 18-32-00062, Deutsches Zentrum fur Luft- und Raumfahrt (DLR), the Swedish National Space Board, and the Swedish Research Council. The Italian authors acknowledge the partial financial support from The Italian Space Agency (ASI) under the program ``Programma PAMELA - attivita' scientifica di analisi dati in fase E''.


\begin{thebibliography}{}

%% Science
\bibitem[\protect\citename{{Adriani} {\em et~al.\ }\relax,  }2011]{2011Sci...332...69A}
Adriani, O., Barbarino, G. C., Bazilevskaya, G. A., et al. 2011, Science, 332, 69
%% Proton solar modulation paper
\bibitem[\protect\citename{{Adriani} {\em et~al.\ }\relax, }2013a]{2013ApJ...765..91}
Adriani, O., Barbarino, G. C., Bazilevskaya, G. A., et al. 2013a, \apj, 765, 91 
%% first H & He isotope paper, ToF only
\bibitem[\protect\citename{{Adriani} {\em et~al.\ }\relax, }2013b]{2013ApJ...770..2}
Adriani, O., Barbarino, G. C., Bazilevskaya, G. A., et al. 2013b, \apj, 770, 2 
%% B & C paper
\bibitem[\protect\citename{{Adriani} {\em et~al.\ }\relax, }2014]{2014ApJ...791..93}
Adriani, O., Barbarino, G. C., Bazilevskaya, G. A., et al. 2014, \apj, 791, 93 
%% second H & He isotope paper with ToF and calorimeter
\bibitem[\protect\citename{{Adriani} {\em et~al.\ }\relax, }2016]{2016ApJ...818..1}
Adriani, O., Barbarino, G. C., Bazilevskaya, G. A., et al. 2016, \apj, 818, 1 

\bibitem[\protect\citename{{Adriani} {\em et~al.\ }\relax, }2017]{NuovoCimento2017}
Adriani, O., Barbarino, G. C., Bazilevskaya, G. A., et al. 2017, RIVISTA DEL NUOVO CIMENTO, 40, 473

\bibitem[\protect\citename{{Agostinelli} {\em et~al.\ }\relax, }2003]{Geant4}
Agostinelli, S., Allison, J., Amako, K., et al. 2003, Nucl. Instrum. Meth. A, 506, 250

\bibitem[\protect\citename{{Ahlen} {\em et~al.\ }\relax, }2000]{2000ApJ...534..757A}
Ahlen, S. P., Greene, N. R., Loomba, D., et al. 2000, \apj, 534, 757  %SMILI

\bibitem[\protect\citename{{Aguilar} {\em et~al.\ }\relax, }2011]{2011ApJ...736..105A}
Aguilar, M., Alcaraz, J., Allaby, J., et al. 2011, \apj, 736, 105  %% Li7/Li6, Be7/(Be9+Be10)

\bibitem[\protect\citename{{Battistoni} {\em et~al.\ }\relax, }2007]{battistoni}
Battistoni, G., Muraro, S., Sala, P. R., et al. 2007, in Hadronic Simulation Workshop 2006,
ed. M. Albrow \& R. Raja, 31


\bibitem[\protect\citename{{Boezio} {\em et al.\ }\relax, }2002]{2002NIMPA.487..407B} 
Boezio, M., Bonvicini, V., Mocchiutti, E., et al. 2002, Nucl. Instr. and Meth. A, 487, 407-422 


\bibitem[\protect\citename{{Brun \& Rademakers} \relax, }1997]{root}
Brun, R. \& Rademakers, F. 1997, ROOT: an object oriented data analysis framework, Nucl.
Instrum. Meth. A 389, 81


\bibitem[\protect\citename{{Bruno}, }2008]{bru08}
%Bruno, G. 1995,  Nucl. Instrum. Meth., A362, 487
Bruno, A., Cosmic ray antiprotons measured in the pamela experiment, Ph.D. thesis, University of Bari, Bari, Italy

\bibitem[\protect\citename{{Buffington} {\em et~al.\ }\relax, }1978]{1978ApJ...226..355B}
Buffington, A., Orth, C. D., Mast, T. S. 1978, \apj, 226, 355  %% UCB balloon Lithium (Beryllium not used in this paper)

\bibitem[\protect\citename{{Connell}\relax, }1998]{1998ApJ...501L..59C}
Connell, J. J. 1998, \apjl, 501, L59  %% Ulysses Be

\bibitem[\protect\citename{{D'Agostini}, }1995]{dagostini}
D'Agostini, G. 1995,  Nucl. Instrum. Meth., A362, 487
%\newblock {A Multidimensional unfolding method based on Bayes' theorem}.

\bibitem[\protect\citename{{de Nolfo} {\em et~al.\ }\relax, }2006]{ACE2006}
de Nolfo, G. A., Moskalenko, I. V., Binns, W. R, et al. 2006 , Adv. Space Res., 38, 1558-1564 % ACE Li6/Li7 also Be7/Be9

\bibitem[\protect\citename{{Garcia-Munoz} {\em et~al.\ }\relax, }1975]{1975ApJ...201..L145}
Garcia-Munoz, M., Mason, G. M., Simpson, J. A. 1975 \apj, 201, L145  %% IMP8 Lithium

\bibitem[\protect\citename{{Garcia-Munoz} {\em et~al.\ }\relax, }1977]{1977ApJ...217..859G}
Garcia-Munoz, M., Mason, G. M., Simpson, J. A. 1977 \apj, 217, 859  %% IMP78 Beryllium

\bibitem[\protect\citename{{Hagen} {\em et~al.\ }\relax, }1977]{1977ApJ...212..262H}
Hagen, F. A., Fisher, A. J., \& Ormes, J. F. 1977, \apj, 212, 262  %%(GSFC Be

\bibitem[\protect\citename{{Hams} {\em et~al.\ }\relax, }2004]{2004ApJ...611..892H}
Hams, T., Barbier, L. M., Bremerich, M., et al. 2004 \apj, 611, 892%% ISOMAX

\bibitem[\protect\citename{{Koldobskiy} {\em et~al.\ }\relax, }2018]{Koldobskiy2018}
Koldobskiy, S. A., Kovaltsov, G. A. \& Usoskin, I. G. 2018, accepted for publication in AGU, doi: 10.1029/2018JA025516

\bibitem[\protect\citename{{MacMillan} \& {Maus}, }2005]{2005EP&S...57.1135M}
{MacMillan}, S., \& {Maus}, S. 2005, Earth, Planets, and Space, 57, 1135

\bibitem[\protect\citename{{Moskalenko \& Mashnik} \relax, }2003]{2003ICRC....4.1969M}
Moskalenko, I.V., \& Mashnik, S.G. 2003, in The 28th Int. Cosmic Ray Conf. (Tsukuba), 2, 1969 


\bibitem[\protect\citename{{Picozza} {\em et~al.\ }\relax, }2007]{2007APh....27..296P}
Picozza, P., Galper, A. M., Castellini, G., et al. 2007, Astroparticle Physics, 27, 296
%\newblock {PAMELA - A payload for antimatter matter exploration and light-nuclei astrophysics}.

\bibitem[\protect\citename{{Reeves} \relax, }1994]{Reeves1994}
Reeves, H., 1994, Rev. Mod. Phys., 66, 193

\bibitem[\protect\citename{{Shea} {\em et~al.\ }\relax, }1987]{1987PEPI...48..200S}
Shea M. A., Smart, D. F., \& Gentile, L. C. 1987, Physics of the Earth and Planetary Interiors, 48, 200--205


\bibitem[\protect\citename{{Strong} \& {Moskalenko}, }1998]{1998ApJ...509..212S}
Strong, A. W., \& Moskalenko, I. V. 1998, \apj, 509, 212 

\bibitem[\protect\citename{{Strong} {\em et~al.\ }\relax, }2007]{2007ARNPS..57..285S}
Strong, A.W., Moskalenko, I. V., \& Ptuskin, V. S. 2007, Annu. Rev. Nucl. Part.
Syst., 57, 285

\bibitem[\protect\citename{{Tomassetti} \relax, }2012]{Tomassetti2012}
Tomassetti, N., N. 2012, Astrophys. Space Sci., 342, 131

\bibitem[\protect\citename{{Vladimirov} {\em et~al.\ }\relax, }2011]{2011CoPhC.182.1156V}
Vladimirov, A. E., Digel, S. W., Jóhannesson, G., et al. 2011 , Comput. Phys. Commun., 182, 1156
	

\bibitem[\protect\citename{{Webber} {\em et~al.\ }\relax, }2002]{2002ApJ...568..210W}
Webber, W. R., Lukasiak, A., McDonald, F. B. 2002 \apj, 568, 210  %% Voyager Li and Be

\bibitem[\protect\citename{{Webber} \relax, }1997]{Webber1997}
Webber, W. R. 1997 Adv. Space Res., 19, 5, 755-758  %% Voyager Li and Be

 
\bibitem[\protect\citename{{Webber \& Kish} \relax, }1979]{NH-balloon}
W. R. Webber and J. Kish 1979, in The 16th Int. Cosmic Ray Conf. (Kyoto), 1, 389   % NH Balloon

\bibitem[\protect\citename{{Wiedenbeck} \& {Greiner}, }1980]{1980ApJ...239L.139W}
Wiedenbeck, M. E., \& Greiner, D. E. 1980, \apj, 239, L139 %%(ISEE 3)






\end{thebibliography}
\end{document}